\documentclass[acmtog]{acmart}
\usepackage{hyperref}
\usepackage[abbreviations]{glossaries-extra}
\usepackage{xspace}
\PassOptionsToPackage{table}{xcolor}
\usepackage{colortbl}
\usepackage{array}
\usepackage{xcolor} 
\usepackage{threeparttable}
\usepackage{subcaption}
\usepackage{afterpage}
\usepackage{lipsum}
\usepackage{booktabs}
\usepackage{algorithm}
\usepackage{algpseudocode}
\usepackage{multirow}
\usepackage{wrapfig}
\usepackage{caption}
\usepackage{amsmath}
\usepackage{sidecap}
\usepackage{xr}
\usepackage{enumitem}
\usepackage{rotating}
\usepackage{pdflscape}
\definecolor{lightred}{rgb}{1,0.4,0.4}
\definecolor{lightgreen}{rgb}{0.6,1,0.3}
\definecolor{forestgreen}{rgb}{0.133, 0.545, 0.133}
\definecolor{lightyellow}{rgb}{1,1.0,0.6}

\setcitestyle{authoryear,round} 
\AtBeginDocument{%
  }

\copyrightyear{2024}
\acmYear{2024}
\acmDOI{XXXXXXX.XXXXXXX}

\acmSubmissionID{Review}

\newcommand{\etal}{~et al.\@\xspace}

\newcommand{\eg}{e.g.\@\xspace}

\newcommand{\refSec}[1]{Sec.~\ref{sec:#1}}
\newcommand{\refSupSec}[1]{Sec.~\ref{supplementary:#1}}
\newcommand{\refFig}[1]{Fig.~\ref{fig:#1}}

\newcommand{\refEq}[1]{Eq.~(\ref{eq:#1})}
\newcommand{\refTbl}[1]{Table~\ref{tbl:#1}}

\newabbreviation{HVS}{HVS}{Human Visual System}
\newabbreviation{AR}{AR}{Augmented Reality}
\newabbreviation{VR}{VR}{Virtual Reality}
\newabbreviation{SLM}{SLM}{Spatial Light Modulator}
\newabbreviation{FoV}{FoV}{Field Of View}
\newabbreviation{HOE}{HOE}{Holographic Optical Element}
\newabbreviation{CNN}{CNN}{Convolutional Neural Network}
\newabbreviation{PSF}{PSF}{Point-Spread Function}
\newabbreviation{MLP}{MLP}{Multilayer Perceptron}
\newabbreviation{CBAM}{CBAM}{Convolutional Block Attention Module}
\newabbreviation{FPN}{FPN}{Feature Pyramid Network}
\newabbreviation{MDE}{MDE}{Monocular Depth Estimation}
\newabbreviation{PSP}{PSP}{Pyramid Spatial Pooling}
\newabbreviation{STE}{STE}{Straight-Through Estimator}
\newabbreviation{BCP}{BCP}{Bin Center Predictor}
\newabbreviation{CGH}{CGH}{Computer-Generated Holography}
\newabbreviation{SA}{SA}{Segment Anything}
\newabbreviation{HDR}{HDR}{High Dynamic Range}
\newabbreviation{LR}{LR}{Learning Rate}
\newabbreviation{TV}{TV}{Total Variation}
\newabbreviation{SILog}{SILog}{Scale Invariant Log}
\newabbreviation{MTL}{MTL}{Multi-task Learning}
\newabbreviation{ASM}{ASM}{Angular Spectrum Method}
\newabbreviation{ViT}{ViT}{Vision Transformer}
\newabbreviation{2D}{2D}{Two-Dimensional}
\newabbreviation{FPS}{FPS}{Frames Per Second}
\newabbreviation{1D}{1D}{One-Dimensional}
\newabbreviation{DP}{DP}{Double Phase}
\newabbreviation{GM}{GM}{Gradient Matching}
\newabbreviation{KD}{KD}{Knowledge Distillation}
\newabbreviation{SAM}{SAM}{Skip Attention Module}
\newabbreviation{SOTA}{SOTA}{state-of-the-art}
\newabbreviation{fp32}{fp32}{32-bit precision}
\newabbreviation{3DGS}{3DGS}{3D Gaussian splatting}
\newabbreviation{NeRF}{NeRF}{neural radiance fields}
\newabbreviation{threeD}{3D}{Three-Dimensional}
\newabbreviation{SH}{SH}{Spherical Harmonic}
\newabbreviation{FFT}{FFT}{Fast Fourier Transform}
\newabbreviation{SBP}{SBP}{Space-Bandwidth Product}
\newabbreviation{DPAC}{DPAC}{Double Phase-Amplitude Coding}
\newabbreviation{OIT}{OIT}{Order Independent Transparency}
\newabbreviation{GWS}{GWS}{Gaussian Wave Splatting}

\global\long\def\GWS{\gls{GWS}\xspace}

\global\long\def\SBP{\gls{SBP}\xspace}
\global\long\def\FFT{\gls{FFT}\xspace}
\global\long\def\STE{\gls{STE}\xspace}
\global\long\def\SH{\gls{SH}\xspace}

\global\long\def\NeRF{\gls{NeRF}\xspace}
\global\long\def\3DGS{\gls{3DGS}\xspace}

\global\long\def\SLM{\gls{SLM}\xspace}

\global\long\def\CGH{\gls{CGH}\xspace}

\global\long\def\ASM{\gls{ASM}\xspace}

\global\long\def\1D{\gls{1D}\xspace}
\global\long\def\2D{\gls{2D}\xspace}

\global\long\def\fp32{\gls{fp32}\xspace}

\newcommand{\PPixel}{P_{i}}

\newcommand{\Pitch}{\Delta x}

\begin{document}

\title{Complex-Valued Holographic Radiance Fields}
\date{\today}

\author{Yicheng Zhan}
\email{UCABY83@ucl.ac.uk}
\affiliation{%
  \institution{University College London}
  \streetaddress{Gower Street}
  \city{London}
  \country{United Kingdom}
  \postcode{WC1E 6BT}
}

\author{Dong-Ha Shin}
\email{0218sdh@gmail.com}
\affiliation{%
  \institution{Pohang University of Science and Technology (POSTECH)}
  \streetaddress{77 Cheongam-ro, Nam-gu}
  \city{Pohang-si}
  \country{South Korea}
  \postcode{37673}
}

\author{Seung-Hwan Baek}
\email{shwbaek@postech.ac.kr}
\affiliation{%
  \institution{Pohang University of Science and Technology (POSTECH)}
  \streetaddress{77 Cheongam-ro, Nam-gu}
  \city{Pohang-si}
  \country{South Korea}
  \postcode{37673}
}

\author{Kaan Akşit}
\affiliation{%
  \institution{University College London}
  \streetaddress{Gower Street}
  \city{London}
  \country{United Kingdom}
}
\email{k.aksit@ucl.ac.uk}

\begin{abstract}
Modeling wave properties of light is an important milestone for advancing physically-based rendering.
In this paper, we propose complex-valued holographic radiance fields, a method that optimizes scenes without relying on intensity-based intermediaries.
By leveraging multi-view images, our method directly optimizes a scene representation using complex-valued Gaussian primitives representing amplitude and phase values aligned with the scene geometry.
Our approach eliminates the need for computationally expensive holographic rendering that typically utilizes a single view of a given scene.
This accelerates holographic rendering speed by 30x-10,000x while achieving on-par image quality with state-of-the-art holography methods,
representing a promising step towards bridging the representation gap between modeling wave properties of light and 3D geometry of scenes.
\end{abstract}



\begin{teaserfigure}
  \centering
  \includegraphics[width=1\linewidth]{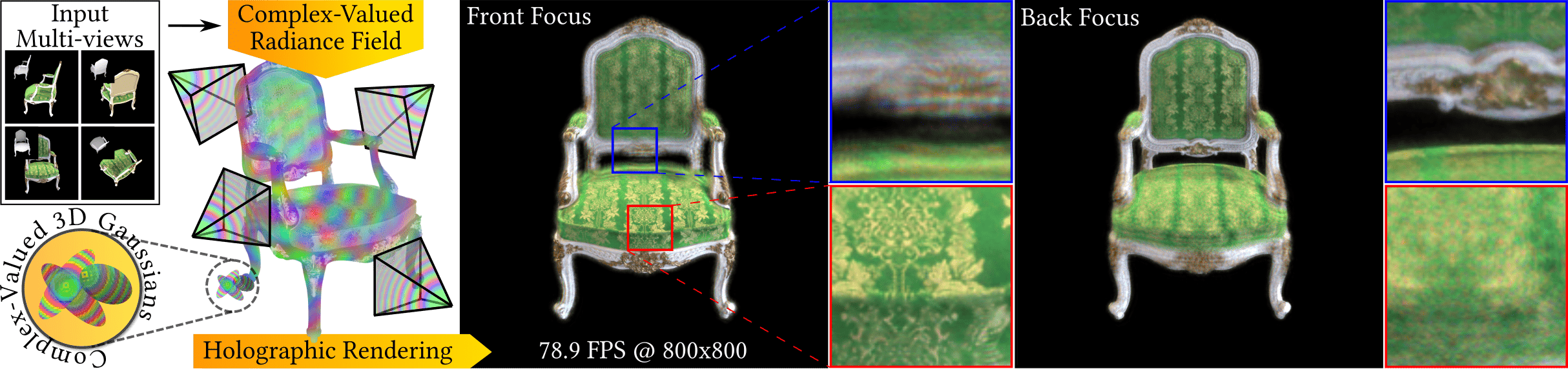}
    \caption{
      Complex-Valued Holographic Radiance Fields.
      Our method explicitly treats the complex-valued holographic radiance fields as the primary optimization target,
      jointly optimizing the 3D scene across multiple viewpoints to model intensity, interference, and diffraction of light in scene representations.
      The complex-valued 3D Gaussians will be projected to multiple depth planes, resulting in a complex 3D hologram via holographic rendering.
      This enables holographic reconstruction with depth-of-field effects: blue boxes highlight the back focus region, and red boxes show the front focus region.
   }
\label{fig:teaser}
\end{teaserfigure}

\maketitle

\section{Introduction}
Light exhibits numerous physical phenomena beyond mere intensity, including wave-optics properties such as spectral variations~\cite{kim2023neural},
polarization~\cite{zhao2022polarimetric}, diffraction, and interference~\cite{goodman2005introduction}.
Accounting for wave-optics properties of light more accurately could help increase the visual fidelity in 3D scene representations~\cite{wen20253d}.
Most recently, \3DGS~\cite{kerbl20233d} has emerged as an effective 3D scene representation to synthesize novel-view intensity images.
However, \3DGS neglects wave-optics properties of light.
In addition, the applications of physically-based rendering and emerging holographic 3D displays~\cite{kim2024holographic} in computer graphics demand modeling these wave-optics properties to be accounted for.

\begin{figure}[t!]
   \centering
   \includegraphics[width=0.47\textwidth]{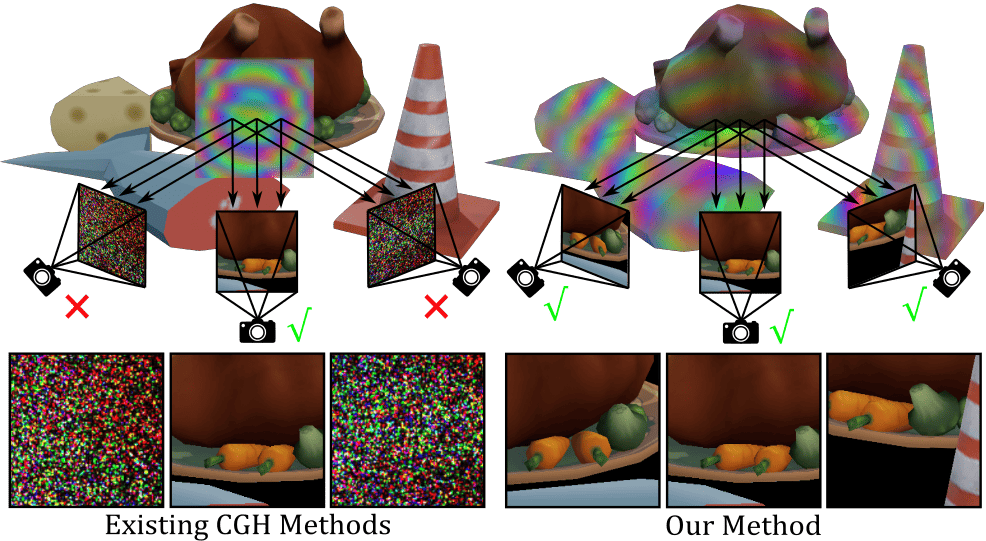}
   \caption{Left: Existing \CGH methods do not preserve geometry of scenes under viewpoint changes, necessitating transformation or recalculation per-view.
   Right: Our complex-valued holographic radiance field offers a 3D consistent representation in contrast.}
   \label{fig:consistency_figure}
\end{figure}

\CGH synthesizes holograms that faithfully reconstruct wavefronts emitted from 3D scenes.
While \CGH has been predominantly utilized for holographic display applications, it has recently garnered increasing interest from the rendering community for generating physically-plausible scenes in computer graphics~\citep{steinberg2023a, steinberg2023b}.
Existing \CGH methods treat hologram synthesis as a post-processing task applied after the conventional rendering pipeline.
These methods typically estimate holograms that reproduce the desired reconstructed intensity image for a fixed viewpoint~\cite{shi2021towards} or a limited viewing angle~\cite{chakravarthula2022pupil, chen2025view}.
When such methods are queried with a new viewpoint, they have to be recalculated for a new hologram using only the provided viewpoint information.
Most recently, the work by Choi~\etal~\shortcite{choi2025gaussian} generates holograms of 3D scenes represented with real-valued, intensity-based 2D Gaussian primitives.
Their method introduces multi-view based approaches tailored for \CGH for the first time, requiring an additional step of transformation for each queried viewpoint, greatly increasing computational complexity.
These methods adopt an \textit{Eulerian} paradigm, in which the hologram plane remains fixed while the complex field is recomputed at each spatial coordinate.
In this context, \textbf{previous CGH methods lack persistent geometric representations,  computing holograms at the camera plane rather than coupling them to 3D scene geometry}.
Thus, such holographic scene representations without any transformation or recalculation would lead to noise-like visuals rather than perspective images when the viewpoint changes, as demonstrated in~\refFig{consistency_figure}.
\setlength{\intextsep}{0.5pt}
\setlength{\columnsep}{10pt}

Our work aims to merge \3DGS radiance fields and holographic rendering under one umbrella using complex numbers,
which are widely adopted in holography for representing both amplitude and phase information from 3D scenes~\cite{goodman2005introduction}.
In this paper, we firstly introduce complex-valued Gaussian primitives.
Our scene representation framework utilizes these complex-valued Gaussian primitives to model both amplitude and phase emitted from 3D scenes.
Unlike previous intensity-based modeling methods~\cite{choi2025gaussian}, our method explicitly treats complex-valued holographic radiance fields as the primary optimization target,
without the need for per-view transformation~\cite{choi2025gaussian} or fixed viewpoint hologram recalculation~\cite{chen2025view}.
Secondly, in our framework the simulation of diffraction and interference phenomena is formulated as a differentiable rasterizer, rendering complex-valued 3D Gaussian primitives from novel views.
In contrast to the existing \CGH methods~\cite{choi2025gaussian} that scale quadratically with the number of primitives, our approach achieves a linear relationship between the primitives and complexity,
achieving 30x-10,000x speed improvements.
By optimizing directly in the complex-valued domain across multiple viewpoints, our method achieves view-dependent diffraction and interference effects without relying on intensity-based intermediaries.
Specifically, our contributions include:
\begin{itemize}
\item Complex-valued holographic radiance fields.
We present a new variant of \3DGS, utilizing complex-valued Gaussian primitives.
Uniquely, our method is transformation-free and follows the geometry of 3D scenes in terms of the phase and amplitude, regardless of the viewpoints.
\item Complex-valued differentiable wave-optics rasterizer.
We enable efficient rendering and optimization of complex-valued holographic radiance fields by rasterizing optical waves emitted from a 3D scene.
\end{itemize}
We evaluate the image quality and speed of our method by comparing against existing Gaussian-based \CGH approaches.
Our codebase is publicly available at our project page~\cite{zhan2025cvhrf_code}.

\section{Related Work}
Our approach bridges recent advances in neural rendering with holography, enabling the efficient generation of holograms from novel views.
We will first review novel-view synthesis methods and then examine existing wave optics rendering techniques that also attempt to simulate light phenomena beyond intensity.
Finally, we discuss the traditional \CGH methods in general, including point-based, polygon-based, and layer-based methods.

\subsection{Novel-View Synthesis}
Traditional novel-view synthesis methods primarily model intensity and appearance, limiting their utility for emerging holographic display
technologies that require phase-aware representations. Early techniques such as image-based rendering~\cite{chaurasia2013depth}
and volumetric scene representations~\cite{hedman2017casual,riegler2020free} laid the groundwork for view-dependent appearance modeling.
Building on these foundations, \NeRF~\cite{mildenhall2021nerf} advanced the field by learning continuous radiance fields that capture view-dependent color and opacity.
This representation has since been extended to improve visual quality and computational efficiency~\cite{barron2021mip,barron2022mip, barron2023zip, chenTensorf2022, Rebain_2021_CVPR, niemeyer2024radsplat, duckworth2024smerf}.

Point-based representations have recently advanced with \3DGS \cite{kerbl20233d}, emerging as a promising alternative to \NeRF.
\3DGS offers a promising foundation for our work, as it encodes both geometry and appearance using oriented Gaussian primitives,
but it is still fundamentally limited to intensity-based representations.
Our work extends \3DGS to incorporate intrinsic amplitude and phase information,
enabling it to model complex-valued holographic radiance fields that capture not only intensity but also interference and diffraction phenomena.

\paragraph{Rendering Beyond Intensity. }
The work by Kim~\etal~\shortcite{kim2023neural} extends \NeRF beyond intensity-based representations to capture spectral and polarimetric properties of light,
revealing richer material and structural characteristics of 3D scenes.
Notably, spectral and polarization data can be supervised using ground-truth measurements obtained from specialized imaging devices.
On the other hand, interference and diffraction phenomena arise from the superposition of phase and amplitude values.
Interference and diffraction effects lack direct measurement data and must be inferred through computational means,
as holographic cameras capable of capturing both amplitude and phase information are not yet commodity equipment~\cite{an2020slim}.
Moreover, supporting simulations of holography in rendering pipelines is a common theme in research.
For instance, Steinberg~\etal~\shortcite{steinberg2023a,steinberg2023b} introduced wave optics formulations for path tracing, enabling partial coherence and diffraction effects.
Choi~\etal~\shortcite{choi2025gaussian} proposed \GWS that transforms 2D Gaussian Splatting~\cite{huang20242d} scenes into holograms by deriving closed-form solutions for angular spectrum.
Chen~\etal~\shortcite{chen2025view} proposed a two-stage method that combines pretrained intensity-based \3DGS with U-Net~\cite{ronneberger2015u}, enabling efficient hologram synthesis for novel views.
In this paper, we propose to enable recalculation-free novel-view hologram synthesis,
whereas Choi~\etal requires per-view wave re-computation and Chen~\etal requires per-view U-Net inference.

\subsection{Eulerian-based Computer-Generated Holography}
Holographic displays rely on \CGH, which algorithmically synthesizes diffraction and interference patterns to reconstruct 3D scenes with near-accurate depth cues.
\textit{Conceptually, traditional \CGH methods follow a per-view Eulerian-based approach to wave propagation},
where the coordinates on the hologram plane remain fixed, while the complex field varies when the relationship between the scene geometry and the viewpoint changes, requiring expensive recalculation.
The computed holographic patterns are implemented using \SLM, which modulates the phase of incident coherent beams to generate designed image patterns.
Through numerical wave propagation simulations~\cite{kavakli2022learned, peng2020neural}, \CGH generates holograms leading to high spatio-angular resolution, positioning it as a promising emerging screen technology.
The accuracy and computational complexity of \CGH strongly depend on primitive representations, which are generally categorized into point-based, polygon-based, and layer-based methods.

Point-based \CGH represents scenes using discrete points emitting spherical waves.
The wave field at the hologram, $P$, is calculated by summing waves from N points,
\begin{equation}
P=\sum_{n=1}^{N}A_n \frac{e^{jkr_n}}{r_n},
\label{eq:pointbased}
\end{equation}
where $A_n$ denotes amplitude, $k$ is the wave number, and $r_n$ is the propagation distance from point $n$ to $P(x,y)$.
Reflecting the Huygens-Fresnel principle, \refEq{pointbased} is highly flexible and physically accurate.
Maimone~\etal~\shortcite{maimone2017holographic} presented point-source-based \CGH computation for holographic near-eye displays for virtual and augmented reality.
Shi \etal~\shortcite{shi2021towards} leveraged point clouds and triangle-ray intersection for occlusion handling in large-scale hologram datasets.
However, the computational cost of point-based \CGH increases significantly with point density, and occlusion handling introduces additional overhead.

Polygon-based \CGH balances efficiency and accuracy by representing surfaces using triangular or polygonal facets emitting coherent wave fields.
This method naturally integrates shading models and intrinsic occlusion handling through visibility algorithms (\eg back-face culling and affine transformations)~\cite{zhang2018fast, matsushima2009extremely}.
Mesh-based \CGH includes topology information for precise occlusion handling~\cite{ahrenberg2008computer, yeom2022efficient}.
While achieving realistic shading by combining wave optics with rendering pipelines, these methods suffer from increased algorithmic complexity and struggle with finely-detailed or non-polygonal structures.

Layer-based \CGH subdivides scenes into discrete depth planes, propagating complex fields from $P$ independently to each plane using operators like the \ASM~\cite{matsushima2009band, zhang2020band, Chuanjun2024SigAsia}.
The $P(x,y)$ is optimized so the propagated results across all planes approximate the desired 3D scene.
Layer-based \CGH greatly accelerates computations through \FFT{}-based propagation, making it suitable for faster rendering and is widely adopted in learned \CGH methods~\cite{shi2022end, choi2021neural, peng2020neural, zhan2024Configure, liu20234k}.
Building on this acceleration, learned complex-valued \CGH methods~\cite{zhong2023real, zhang2025real} generate phase-only holograms from complex-valued fields in a single forward pass.
These approaches share the same application as our method for SLM-based holographic displays.
In contrast, our method embeds complex-valued primitives directly into the 3D scene representation, eliminating the need for re-inference per viewpoint.

While recent \CGH advancements enable realistic defocus effects and continuous pupil accommodation~\cite{kim2024holographic, kavakli2023realistic},
existing light field–based 3D and 4D \CGH methods are typically optimized for limited viewing directions, with computed holograms losing the scene information under large viewpoint changes.
We propose integrating strengths of both point-based and layer-based \CGH methods,
utilizing complex-valued 3D Gaussians as point-based primitives to encode scene geometry and complex-valued radiance.
\textit{Unlike existing Eulerian-based \CGH approaches, our method follows a Lagrangian-based formulation}, treating amplitude and phase as intrinsic variables in scene representations.
Compared with point clouds and polygon-meshes, Gaussians have learnable scale and rotation parameters, allowing them to capture anisotropic structures and fine-grained spatial variations effectively.
This expressiveness enables both effective appearance modeling and geometrically faithful wave emitting distribution.
\refTbl{method_comparison} presents a comparison between our method and other Gaussian-based \CGH approaches,
demonstrating its ability to achieve fast inference, natural defocus blur, on-par image quality, and scene geometry-aware amplitude and phase representations across novel views.
\begin{table}[t!]
\centering
\caption{
Comparison of Gaussian primitives-based hologram synthesis method.
\textit{Inference} refers to inference speed.
\textit{Re} refers to whether the method requires recalculation when geometric relationship between scene and observer changes.
\textit{Type} refers to Gaussian representation type.
\textit{Quality} refers to the image quality of reconstructed images from the holograms.
\textit{Align} refers to scene geometry-aware amplitude and phase representations across novel views.
\textit{NDB} refers to natural defocus blur.
}
\label{tbl:method_comparison}
\begin{threeparttable}
\footnotesize
\begin{tabular}{m{2.2cm} m{0.9cm} m{0.35cm} m{0.8cm} m{0.75cm} m{0.65cm} m{0.4cm}}
\toprule
\multirow{2}{0cm}{\centering \textbf{Method}} & \multirow{2}{0cm}{\centering \textbf{Inference}} & \multirow{2}{0cm}{\centering \textbf{Re}} & \multirow{2}{0cm}{\centering \textbf{Type}} & \multirow{2}{0cm}{\centering \textbf{Quality}} & \multirow{2}{0cm}{\centering \textbf{Align}} & \multirow{2}{0cm}{\centering \textbf{NDB}} \\
\addlinespace[0.5em]
\midrule
\renewcommand{\arraystretch}{1.2}
\parbox{2.2cm}{3DGS + U-Net~\shortcite{chen2025view}} & \cellcolor{lightgreen}Fast & \cellcolor{lightred}Yes & \cellcolor{lightred}Intensity & \cellcolor{lightgreen}Good & \cellcolor{lightred}No & \cellcolor{lightred}No \\
\parbox{2.2cm}{GWS~\shortcite{choi2025gaussian}} & \cellcolor{lightyellow}Moderate & \cellcolor{lightred}Yes & \cellcolor{lightred}Intensity & \cellcolor{lightgreen}Good & \cellcolor{lightred}No & \cellcolor{lightred}No \\
\parbox{2.2cm}{Our Method} & \cellcolor{lightgreen}Fast & \cellcolor{lightgreen}No & \cellcolor{lightgreen}Complex & \cellcolor{lightyellow}On-par & \cellcolor{lightgreen}Yes & \cellcolor{lightgreen}Yes \\
\bottomrule
\end{tabular}
\end{threeparttable}
\end{table}
\begin{figure*}[t!]
   \centering
   \includegraphics[width=1\textwidth]{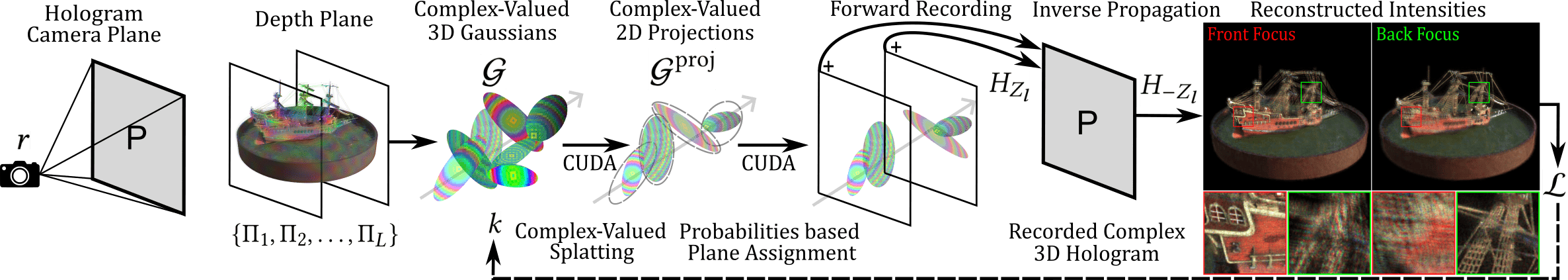}
   \caption{
   Without the need to query an intensity-based radiance field,
   our approach models a complex-valued holographic radiance field using Gaussian primitives with intrinsic amplitude and phase properties,
   enabling scene geometry-aware amplitude and phase modeling across viewpoints and efficient rendering through a differentiable multi-layer propagation pipeline. }
   \label{fig:system}

 \end{figure*}
\section{Background and Problem Definition}
\label{sec:background}
Before explaining our approach, we first introduce the key concepts and terminology in this paper to define the research problem we are trying to solve.
These concepts are: 3D Gaussians for scene representations, hologram planes for wave recording, and camera parameters for view-dependent rendering.

\paragraph{Representing 3D scenes with intensity-based Gaussians.}
In line with the approach of Kerbl \etal~\shortcite{kerbl20233d}, we learn geometry and radiance of scenes through the optimization of 3D Gaussian primitives.
We consider a 3D region in Euclidean space, $\Omega \subset \mathbb{R}^{3}$.
A scene inside $\Omega$ is described using a set of $N$ oriented 3D Gaussians, and each Gaussian is denoted by $\mathcal{G}_n$.
Typically, each $\mathcal{G}_n$ is parameterized by $k = (\mathbf{c}_n, \mathbf{x}_n, \mathbf{R}_n, \mathbf{S}_n, \boldsymbol{\alpha}_n)$,
where $\mathbf{c}_n \in \mathbb{R}^{3}$ represents color, $\mathbf{x}_n \in \mathbb{R}^3$ represents the center point of a Gaussian,
$\mathbf{R}_n \in \mathbb{R}^{4}$ describes the rotation of a Gaussian,
$\mathbf{S}_n$ represents scale along principal axes defined by $\mathbf{R}_n$,
and $\boldsymbol{\alpha}_n$ represents the opacity.
The \emph{geometrical shape} of a 3D Gaussian is defined as
\begin{equation}
 \mathcal{G}(\mathbf{x}, \mathbf{R}, \mathbf{S}) \;=\; \exp \Bigl(-\tfrac{1}{2}\,\mathbf{x}^{\top}\,\Sigma^{-1}\,\mathbf{x}\Bigr),
   \label{eq:gaussian_density}
\end{equation}
where the covariance, $\Sigma$, is decomposed into a rotation matrix $\mathbf{R}$, and scaling matrix $\mathbf{S}$, following
\begin{equation}
 \Sigma \;=\; \mathbf{R}\,\mathbf{S}\,\mathbf{S}^{\top}\,\mathbf{R}^{\top}.
   \label{eq:gaussian_decompose}
\end{equation}
We project these 3D Gaussians onto 2D planes to render images at a camera plane, a process also known as the \emph{differential splatting}.
The process of differential splatting involves applying the transformation matrix, $\mathbf{W}$, and the Jacobian projection function, $\mathbf{J}$,
mapping 3D points to 2D image plane.
The covariance in camera-space, $\Sigma'$, is defined as
\begin{equation}
   \Sigma' \;=\; \mathbf{J}\,\mathbf{W}\,\Sigma\,\mathbf{W}^{\top}\,\mathbf{J}^{\top}.
   \label{eq:gaussian_projection}
\end{equation}
For each pixel, the color and opacity contributed by each Gaussian
are computed by evaluating \refEq{gaussian_density} under its learned parameters.
The blending of $N$ ordered Gaussians overlapping a given pixel is defined as follows
\begin{equation}
   \mathbf{C}_{N} \;=\;
   \sum_{n=1}^{N} \,
   \mathbf{c}_{n}\,\boldsymbol{\alpha}_{n}
   \prod_{j=1}^{n-1}\!\bigl(1 - \boldsymbol{\alpha}_{j}\bigr),
   \label{eq:gaussian_blending}
\end{equation}
where $\mathbf{c}_{i}$ and $\boldsymbol{\alpha}_{i}$ represent the color and opacity of the $i$-th Gaussian.

\paragraph{Hologram Plane, Camera and Problem Definition.}
We define the hologram plane as $P \in \mathbb{C}^{n_x \times n_y}$, where $n_x, n_y \in \mathbb{Z}$.
The center of $P$ is co-located with a camera pointing at the center of $\Omega$ located at $d$ distance away.
We always position $P$ based on the camera pose, location and angles, $r = (x_c, y_c, z_c, \phi_x, \phi_y, \phi_z)$, in Euler space.
Each complex pixel $\PPixel$ has a pixel pitch, $\Pitch$.
Given input $q = (r, \Pitch, n_x, n_y)$, we compute the hologram plane by propagating the complex field rendered from Gaussian primitives to $P(q)$,
\begin{equation}
P(q) = U(q) * h,
\end{equation}
where $U(q)$ denotes the complex field rendered from the Gaussian primitives under camera parameters $q$,
and $h$ denotes spatial-domain convolutional kernel~\cite{matsushima2009band, Chuanjun2024SigAsia}.
Note that this propagation is an intermediate, differentiable rendering step.
Our final goal is to optimize the Gaussian parameters such that the reconstructed intensities after inverse propagation match the focal-stack.
A common propagation method is the band-limited \ASM, where the transfer function $H_{z}(f_x, f_y)$
encodes propagation by a distance $z$ in the frequency domain and its inverse Fourier transform yields the spatial-domain kernel $h$,
\begin{equation}
 H_{z}(f_x, f_y) =
 \begin{cases}
 \exp(j2\pi z\sqrt{\tfrac{1}{\lambda^2} - (f_x^2 + f_y^2)}), & \text{if } f_x^2 + f_y^2 \leq \tfrac{1}{\lambda^2} \\
   0, & \text{otherwise}.
 \end{cases}
   \label{eq:band_limited_asm_transfer_function}
\end{equation}
In this paper, we also utilize band-limited \ASM as the propagation method for complex-valued holographic radiance fields modeling.

\section{Method}
\label{sec:method}
In this section, we present our method using the terminology established in \refSec{background}.
Whereas classical radiance is a scalar energy measure $\mathrm{W/(sr\cdot m^2)}$ neglecting phase,
our approach represents a complex-valued field, capturing both amplitude and phase information to model intensity, interference and diffraction.
We use the term \textit{complex-valued holographic radiance fields}, preserving conceptual links to prior 3D scene representations
such as \NeRF\ and \3DGS while reinterpreting radiance within the wave-optics framework.
By embedding coherent light modeling into the established radiance field paradigm, our formulation attempts to bridge graphics and holographic rendering.

\subsection{Complex-Valued Holographic Radiance Fields}
\label{sec:view_dependent_layer}
We provide an overview of our method in \refFig{system}.
Given a camera pose, $r$, and a target hologram plane, $P$,
we partition $\Omega$ into multiple planes in depth.
At each depth plane, we render our learnable complex-valued 3D Gaussians into corresponding 2D projections.
During projection, we utilize learnable probabilities for our 3D Gaussians that help us determine if a 3D Gaussian contributes to that selected depth plane.
Note that learning these probabilities in terms of their contributions at each depth plane helps to inherently capture depth characteristics of a given scene.
Once we determine the appropriate 2D projections for each layer,
we propagate the complex field from each layer towards our camera to construct our $P$.

\paragraph{Layer Construction.}
For every 3D Gaussian $\mathcal{G}_n$ with center $\mathbf{x}_n$,
we project its 2D projections to plane $\Pi_{l}$ based on learned assignment probabilities.
Let $\{\Pi_1, \Pi_2, \dots, \Pi_L\}$ be a set of discretized parallel planes in the volume, $\Omega$,
positioned periodically along the optical axis in front of the view-dependent camera.
Each plane $\Pi_l$ is separated from its neighbors by a fixed depth interval of $\Delta z$.
The planes span a volume of depth and the center of the volume is placed at a propagation distance, $d$, from the camera plane.
In this arrangement, the entire depth is symmetrically distributed as in the common literature, ranging from:
$Z_1 = d - \frac{(L-1)}{2} \Delta z$ to $Z_L = d + \frac{(L-1)}{2} \Delta z.$
Each plane $\Pi_l$ is sampled on a 2D grid that matches the physical resolution of a \SLM.

\paragraph{Complex-Valued 3D Gaussian.}
Consider a multi-plane case where Gaussians are projected onto $L$ numbers of planes $\{\Pi_l\}_{l=1}^{L}$,
we compute each 2D projection of 3D Gaussian following \refEq{gaussian_projection},
\begin{equation}
 \mathcal{G}^{\text{proj}}(\mathbf{x}, \mathbf{R}, \mathbf{S}, \mathbf{J}, \mathbf{W}, u) = \exp\left(-\tfrac{1}{2} (u - \mu)^{\top} \Sigma'^{-1} (u - \mu)\right),
    \label{eq:gaussian_projection_2d}
\end{equation}
where $u = (n_x, n_y)$ and $\mu = \mathbf{J}\,\mathbf{W}\,\mathbf{x}$.
To model complex-valued holographic radiance fields, we have to modify the original definition of the 3D Gaussian primitive.
\textbf{Specifically, each Gaussian primitive is now parameterized by }
\begin{equation}
k = (\mathbf{c}_n, \mathbf{x}_n, \mathbf{R}_n, \mathbf{S}_n, \boldsymbol{\alpha}_n, \underline{\boldsymbol{\varphi}_n}, \underline{\boldsymbol{\rho}_n}),
\end{equation}
where we add two extra parameters: $\boldsymbol{\varphi}_n \in \mathbb{R}^3$ to represent inherent phase values across wavelengths,
and $\boldsymbol{\rho}_n \in \mathbb{R}^{L}$ to represent the plane assignment probabilities, where each element lies in $[0, 1]$.
Under this formulation, $\mathbf{c}_n$ denotes the inherent wave amplitude, replacing its original interpretation as color.
The complex field of each 2D projection $U_n$ then is given by
\begin{equation}
U_n = \mathbf{c}_n \mathcal{G}_{n}^{\text{proj}} \exp(j \boldsymbol{\varphi}_n).
\label{eq:single_plane_field}
\end{equation}
The wavenumber term is implicitly included in the phase parameter $\boldsymbol{\varphi}_n$, which represents the accumulated phase (including wavelength dependence) at the Gaussian level.
When $L = 1$, plane assignment reduces to a single plane, the $\boldsymbol{\rho}_n$ trivially equals 1.0, and the complex-valued holographic radiance field will generate 2D holograms.
For the multi-plane case, we convert the raw, pre-activation logits $\boldsymbol{\rho}'_n$ to actual probabilities via arg-max followed by one-hot encoding,
resulting in a hard assignment vector $\boldsymbol{\rho}_n$
\begin{equation}
\boldsymbol{\rho}_{n,l} = \mathrm{OneHot}(\arg\max_{l}\,\boldsymbol{\rho}'_{n}) =
\begin{cases}
1, & \text{if } l = \displaystyle\arg\max_{l}\,\boldsymbol{\rho}'_{n}, \\
0, & \text{otherwise},
\end{cases}
\label{eq:hard_assignment}
\end{equation}
where $\boldsymbol{\rho}_{n,l}$ is the assignment probability of the $n$-th Gaussian to the $l$-th depth plane.
However, the argmax operation is non-differentiable, which impedes gradient-based training.
To address this, we employ the \STE that allows gradient propagation through the discrete assignment
\begin{equation}
   \begin{aligned}
   \text{Forward: } & \boldsymbol{\rho}_n = \text{OneHot}(\underset{l}{\arg\max}\, \boldsymbol{\rho}'_{n}), \\
   \text{Backward: } & \frac{\partial \mathcal{L}}{\partial \boldsymbol{\rho}'_n} = \frac{\partial \mathcal{L}}{\partial \boldsymbol{\rho}_n} \cdot \text{softmax}(\boldsymbol{\rho}'_n / \tau),
   \end{aligned}
   \label{eq:ste}
\end{equation}
where $\tau$ is the temperature controlling the sharpness of the softmax approximation, in our case, we set $\tau=0.001$, which makes the softmax function close to argmax operation.
The \STE is widely used in deep learning to approximate gradients through discrete operations~\cite{Yin2019STE}, but it is not a standard component of neural rendering pipelines.
We adopt it to our method to circumvent gradient propagation issues.

\paragraph{Forward Recording and Inverse Propagation.} The final complex field at each plane $\Pi_l$ is then defined as
\begin{equation}
U_{\Pi_{l}} = \sum_{n=1}^{N} \boldsymbol{\rho}_{n,l} U_n \boldsymbol{\alpha}_{n} \prod_{j=1}^{n-1} \bigl(1 - \boldsymbol{\alpha}_{j}\bigr) \boldsymbol{\rho}_{j,l}.
\end{equation}
This formulation ensures each complex-valued 3D Gaussian contributes exclusively to its assigned plane.
Here, the opacity $\alpha_n$ modulates the amplitude of the emitted complex wavefront, representing electromagnetic field strength rather than color blending.
It controls the wave emission intensity of each Gaussian and thereby indirectly shapes the resulting interference pattern,
while the transmittance product $\prod_{j=1}^{n-1}(1-\alpha_j)$ accounts for wavefront attenuation due to occlusion by preceding Gaussians.
The learned $\boldsymbol{\rho}_n$ values guide assignment distribution across view-dependent plane configurations at different viewpoints, minimizing reconstruction loss during training.
Unlike the alpha blending used in intensity-based approaches, which indiscriminately mixes RGB colors across all elements,
our method creates plane-specific visibility by blending complex numbers (containing both amplitude and phase information) only among Gaussians assigned to the same depth plane to each pixel.
Once each layer $\Pi_{l}$ is populated by assigned 2D projections,
we use the transfer function $H$ to propagate $\Pi_{l}$ towards the hologram plane $P$, a process we named \textit{Forward Recording} in our framework.
This wave-based propagation extends the traditional intensity neural rendering in \refEq{gaussian_blending}.
Specifically, for every $\Pi_l$, a $(n_{x}^{(l)}\times n_{y}^{(l)})$ grid of complex samples, we compute
\begin{equation}
   U_{\Pi_{l} \to P} = \mathcal{F}^{-1}\{
      H_{Z_l}(f_x,f_y)
      \cdot
      \mathcal{F}\{U_{\Pi_{l}}\}
   \},
   \label{eq:layer2plane_asm}
\end{equation}
where $Z_l$ is the distance between depth plane $\Pi_l$ and the hologram plane $P$.
The forward propagated field $U_{\Pi_{l} \to P}$ is then added with the contributions from the other planes to record the final complex 3D hologram~\cite{Shimobaba09},
\begin{equation}
   P = \sum_{l=1}^{L} U_{\Pi_{l} \to P}.
   \label{eq:sum_planes}
\end{equation}
Since each plane is a 2D raster, the computation cost scales linearly with the number of layers $L$ and is dominated by \FFT{}s of size $n_x^{(l)} \times n_y^{(l)}$.
After obtaining $P$, we propagate it back to each depth plane.
The back-propagated complex field $U_{ P \to \Pi_{l}}$ at depth plane $\Pi_{l}$ is then computed as
\begin{equation}
   U_{ P \to \Pi_{l}} = \mathcal{F}^{-1}\{
      H_{-Z_l}(f_x,f_y)
      \cdot
      \mathcal{F}\{P\}
   \},
   \label{eq:plane2layer_asm}
\end{equation}
where we obtain its intensity $I_{l} = |U_{P \to \Pi_{l}}|^2$ as the final rendered image, we name this process \textit{Inverse Propagation} in our framework.
Naively, we can supervise the reconstructed intensities $I$ from back-propagated complex field $U$ against the ground-truth image $I_{gt}$ per depth plane $l$
\begin{equation}
   \mathcal{L}_{MSE} = \frac{1}{L} \sum_{l=1}^{L} \|I_l - I_{gt, l}\|^2.
   \label{eq:recon_loss}
\end{equation}
The supervision consists of focal-stack intensity images rendered from captured or synthetic multi-view data~\cite{kavakli2023realistic},
serving as ground-truth observations and are not produced by any learned baseline.
To further improve the image quality of the defocus region, we utilize two complementary loss functions.
First, the reconstruction loss $\mathcal{L}_{recon}$ introduced by Kavakli\etal~\shortcite{kavakli2023realistic}, computed as
\begin{equation}
   \begin{split}
   \mathcal{L}_{recon} &= \frac{1}{L} \sum_{l=1}^{L} ( \|I_l - I_{gt, l}\|^2 \\
   &\quad + \|I_l \cdot M_l - I_{gt, l} \cdot M_l\|^2 + \|I_l \cdot I_{gt, l} - I_{gt, l} \cdot I_{gt, l}\|^2 ),
   \end{split}
\end{equation}
where $M_l$ is the binary mask for depth plane $\Pi_l$ generated from the target image and its quantized depth.
Additionally, we employ the SSIM loss $\mathcal{L}_{SSIM}$ defined as
\begin{equation}
   \mathcal{L}_{SSIM} = \frac{1}{L} \sum_{l=1}^{L} \lambda_{1} \cdot (1 - \text{SSIM}(I_l, I_{gt, l})),
\end{equation}
where $\lambda_{1} = 0.005$.
The final training loss is $\mathcal{L} = \mathcal{L}_{recon} + \mathcal{L}_{SSIM}$.
Neither the complex hologram values nor the plane-assignment probabilities are directly supervised in our method,
as no GT hologram or plane label per Gaussian is available.
Both variables are learned implicitly through reconstruction loss on focal-stack intensities.

\paragraph{View-Dependent Plane Configuration.}
At each viewpoint, the depth planes $\{\Pi_l\}_{l=1}^{L}$ are always positioned parallel to the hologram plane $P$, ensuring standard band-limited \ASM remains valid.
To our knowledge, there is not a work adopting off-axis beam propagation in this context, including the prior work~\cite{choi2025gaussian}.
For novel views with camera pose $r'$, we recompute projection matrices $\mathbf{J}'$ and $\mathbf{W}'$ to project Gaussians onto the reconfigured planes.
The plane assignment probabilities $\boldsymbol{\rho}_n$ are learned across training views, serving as computational latent variables that guide Gaussian distribution across depth planes.

\subsection{Fast Differentiable Complex-Valued Rasterizer}
\label{sec:fast_rast}
Building upon the tile-based architecture from 3DGS, we extend the system to support complex field calculations for holographic rendering while
maintaining fast sorting and efficient parallel processing. Our method preserves the core 16×16 tile structure,
computing screen-space extents with additional consideration for phase-dependent effects during preprocessing.

For the forward pass, we adapt tile-based rendering to track both real and imaginary components of Gaussians.
Each thread block collaboratively loads Gaussian data into shared memory, with threads processing individual pixels by accumulating
complex-valued field contributions using fast trigonometric operations.
Multi-plane rendering is handled through plane-specific filtering directly in the kernel,
skipping Gaussians with low assignment probabilities to minimize thread divergence while supporting depth-dependent complex field accumulation.
Per pixel, we track final transmittance and the last contributing Gaussian's position per plane for correct backward gradients.

The backward pass maintains the same tile-based structure but traverses Gaussians back-to-front,
reusing the sorted array from the forward pass with pixels processing only up to their recorded last contributor.
This ensures constant memory overhead regardless of scene complexity.
We store only the final accumulated opacity per pixel during forward pass,
then recover intermediate opacity values during the backward pass by dividing the final opacity by each Gaussian's contribution,
eliminating the need for long opacity lists while enabling accurate gradient computation.
This implementation efficiently renders complex-valued holographic radiance fields while retaining the performance benefits
of the original tile-based rendering structure. For details of the forward operation and backward gradient computation,
please refer to Supplementary~\refSupSec{cuda_gradient}.

\subsection{Scene Geometry-Aware Amplitude and Phase Representations}
\paragraph{Existing \CGH Methods. }
A limitation in existing \CGH methods is their dependency on optimizing the complex field for a fixed geometric relationship between the scene geometry and the hologram plane.
In these methods, a hologram plane $P_0(x,y)$ is designed to encode the entire 3D scene when viewed from a specific viewpoint,
with the complex field at the viewing plane described as:
\begin{equation}
U_0(x',y') = \mathcal{F}^{-1}\{
   H_{z}(f_x,f_y)
   \cdot
   \mathcal{F}\{P_0(x,y) \cdot R(x,y)\}
\},
\end{equation}
where $R(x,y)$ is the reference wave.
While a properly estimated 3D hologram inherently contains information for different viewing angles within the \SBP limits of the pixel pitch—
functioning as a ``window'' into a virtual world that observers can view from different positions—
the issue arises when the \textit{geometric relationship between the scene geometry and the hologram plane changes}.
Unlike assuming a stationary or narrowly perturbed observer at the center of projection,
any significant movement of the camera plane itself within the 3D world demands recalculating the hologram:
\begin{equation}
P_{\theta}(x,y) \neq P_0(x,y),
\end{equation}
where $\theta$ represents the new viewpoint when the geometric relationship between scene geometry and hologram plane changes.
Existing \CGH methods estimate, optimize, or calculate closed-form solutions of $P(x,y)$
to produce desired intensity patterns, treating each viewpoint as an independent optimization problem.
As shown in ~\refFig{consistency_figure}, when the scene-to-hologram relationship changes, the previously computed hologram leads to visuals that are highly distorted, resembling noise.
Although neural networks enable efficient recalculation of holograms for new configurations,
these learned methods fail to maintain consistent amplitude and phase across different viewpoints.
Instead, neural networks essentially create a new, unrelated data-driven solution,
which lacks interpretability and physical consistency;
in real-world scenarios, the amplitude and phase relationships from an object remain invariant regardless of its geometric relationship to the observation plane.
\begin{figure*}[t!]
   \centering
   \includegraphics[width=0.97\textwidth]{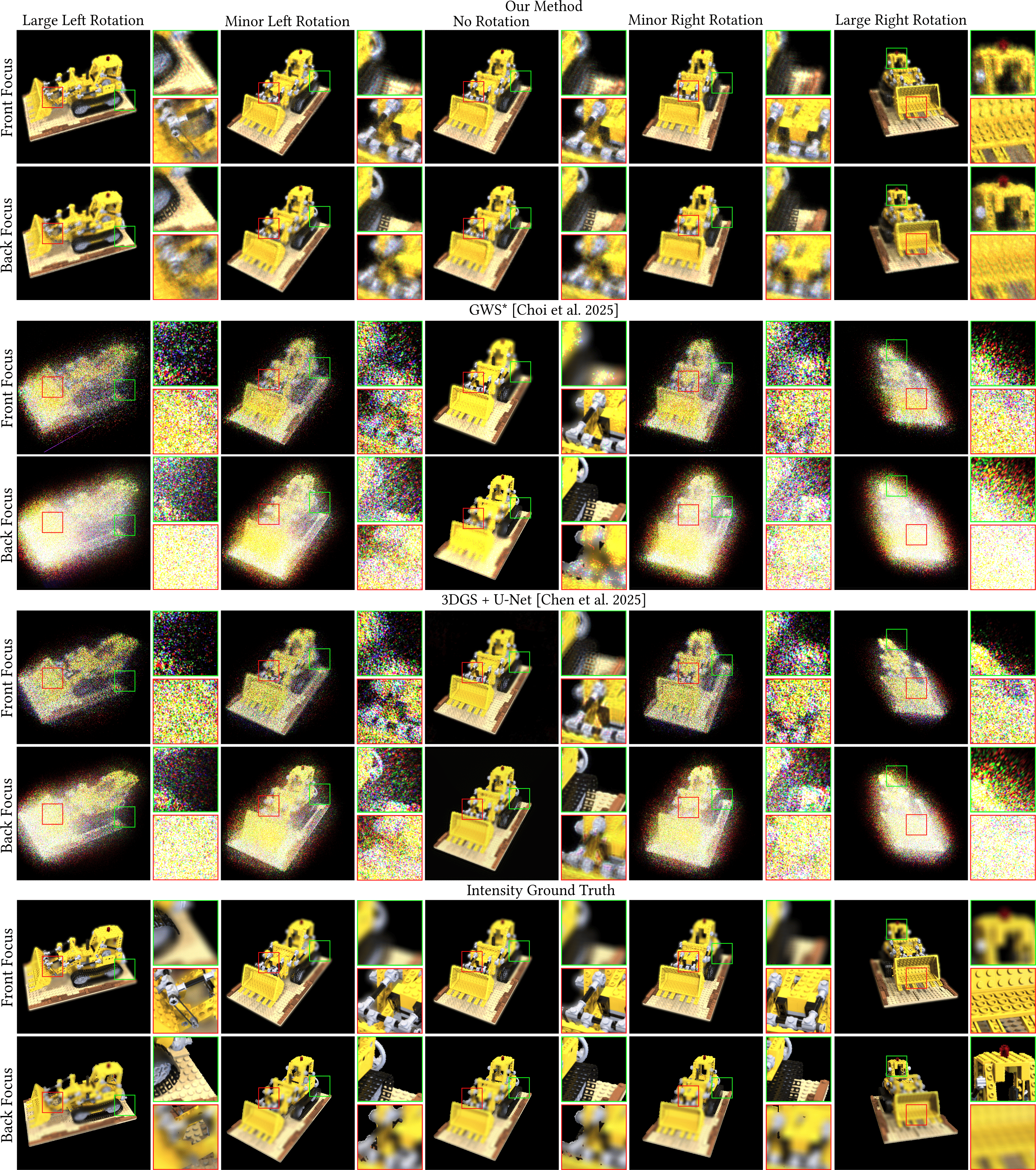}
   \caption{Comparison of different hologram synthesis methods across different viewpoints in simulation.
   The top row shows our method with scene geometry-aware representations from large left to right rotations for novel views.
   The middle rows show the existing \CGH methods that rely on intensity-based intermediaries, which fail to maintain consistency across novel views.
   The bottom row shows the intensity ground truth.
   \textsuperscript{*} We reimplement \cite{choi2025gaussian} to demonstrate the results of \GWS;
   The original non-rotational result shows higher image quality and different defocus blur than our reimplementation.
   }
   \label{fig:phase_consistent}
\end{figure*}
In traditional intensity-based rasterization, each pixel depends only on a handful of nearby fragments; in \CGH, by contrast,
every primitive contributes across a large ``sub-hologram'' footprint on the full resolution hologram,
and one must accumulate complex-valued wavefronts from all primitives at all spatial frequencies.
Primitive-based methods that maintain physical accuracy (e.g., \GWS~\cite{choi2025gaussian} or polygon-based \CGH)
require proper handling of occlusions and interference effects through complex wavefront accumulation:
\begin{equation}
P(x,y) = \sum_{n=1}^{N} \mathcal{F}^{-1}\{
   H_{z_n}(f_x,f_y)
   \cdot
   \mathcal{F}\{\mathcal{G}_n(x,y) \cdot \mathcal{T}_n(x,y)\}
\},
\end{equation}
where $\mathcal{T}_n(x,y)$ represents the accumulated transmittance for occlusion of the $n$-th primitive.
This formulation requires expensive convolution for each primitive across the entire hologram plane,
yielding complexity of $O(N_{\text{primitives}} \times N_{\text{freq}} \log N_{\text{freq}})$, where $N_{\text{freq}}$ is the total number of frequency bins.
To improve efficiency, occlusion effects can be neglected by approximating $\mathcal{T}_n(x,y)\approx1$:
\begin{equation}
P(x,y) \approx \sum_{n=1}^{N} \mathcal{F}^{-1}\{
   H_{z_n}(f_x,f_y)
   \cdot
   \mathcal{F}\{\mathcal{G}_n(x,y)\}
\}.
\end{equation}
While this reduces complexity to $O(N_{\text{primitives}} \times N_{\text{freq}})$ and enables better parallelization,
it sacrifices accurate occlusion modeling and still requires substantial computation, as each primitive must be re-evaluated across the entire hologram plane.
Existing \CGH methods follow an Eulerian computational paradigm, where the hologram plane coordinates remain fixed while complex fields are recalculated for each viewpoint.

\begin{figure*}[t!]
   \centering
   \includegraphics[width=0.99\textwidth]{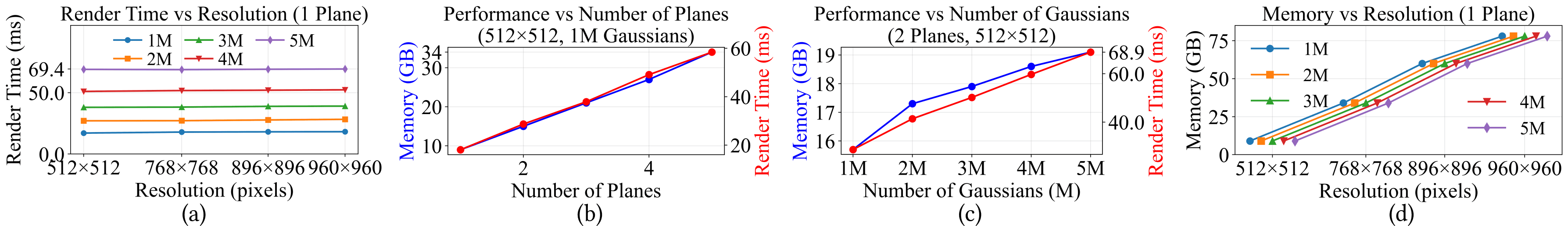}
   \caption{Performance and memory usage analysis.
   (a) Render time vs resolution for different Gaussian counts under 1 plane.
   (b) Memory usage \& render time vs number of depth planes.
   (c) Memory usage \& render time vs number of Gaussians.
   (d) Memory usage vs resolution for different numbers of Gaussians.}
   \label{fig:infer_time}
\end{figure*}

\paragraph{Our Method.}
In contrast, our approach differs fundamentally by modeling the complex field directly in 3D space using Gaussian
primitives with intrinsic phase properties — a Lagrangian perspective where the content remains fixed while the viewpoint changes.
Each complex-valued Gaussian $\mathcal{G}_n$ has an explicitly defined phase parameter $\boldsymbol{\varphi}_n$ that represents its inherent holographic properties,
independent to camera parameters. We emphasize that this per-Gaussian phase parameter does not correspond to the absolute optical phase
measured at the sensor in real life and does not change with the position of the observer or the camera.
Instead, it serves as a learned intrinsic reference phase, characterizing the local wave emission of each Gaussian within the scene`s Lagrangian coordinate.
View-dependent and position-dependent interference will arise subsequently through light propagation.
Although the intrinsic phase parameters remain fixed across viewpoints, the result of interference observed at the camera plane will vary consistently with scene geometry.
Therefore, the complex field contribution of each Gaussian is uniquely determined by its 3D position, orientation, and intrinsic phase, remaining coherent across views.

This Lagrangian formulation enables us to avoid recalculations in the Eulerian approach,
avoids the expensive per-frequency summation, and instead pushes all of the work into two highly optimized GPU-friendly stages:
\textit{1. Tile-based Complex-Valued Rasterization: }
As described in \refSec{fast_rast}, by dividing the screen into 16×16 tiles, we extend the \3DGS rasterizer to accumulate complex-valued Gaussians directly in image space and
reduce the rasterization cost to $O(N_{\text{primitives}})$.
\textit{2. FFT-based layer propagation: }
Instead of summing each primitive's contribution at every spatial frequency, we collect all Gaussians onto $L$ depth planes once,
and then propagate these planes to the hologram plane using 2D FFTs.
This reduces the propagation cost to $O(L \times N_{\text{res}} \log N_{\text{res}})$, where $N_{\text{res}} = n_x \times n_y$ is the number of pixels per depth plane, independent to the number of primitives.
In this case, since $L$ does not grow with $N_{\text{primitives}}$ and $L \times N \log N$ is much smaller
than $N_{\text{primitives}} \times N_{\text{freq}}$ when the number of primitives is large, our approach achieves inherently better scaling.
This results in a more straightforward holographic rendering pipeline that maintains constant propagation cost regardless of scene complexity.
Mathematically, consider two different viewpoints with respective projection matrices $\mathbf{J}_1$, $\mathbf{W}_1$ and $\mathbf{J}_2$, $\mathbf{W}_2$.
For a given $\mathcal{G}_n$, while its projected complex field will differ between viewpoints,
the underlying phase relationship remains consistent:
\begin{equation}
\begin{aligned}
U_n^{(1)} &= \mathbf{c}_n \mathcal{G}_{n}^{\text{proj}}(\mathbf{J}_1, \mathbf{W}_1) \exp(j \boldsymbol{\varphi}_n), \\
U_n^{(2)} &= \mathbf{c}_n \mathcal{G}_{n}^{\text{proj}}(\mathbf{J}_2, \mathbf{W}_2) \exp(j \boldsymbol{\varphi}_n).
\end{aligned}
\end{equation}
This phase consistency is further maintained through the \textit{Forward Recording} process:
\begin{equation}
P^{(i)} = \sum_{l=1}^{L} \mathcal{F}^{-1}\{
   H_{Z_l}(f_x,f_y)
   \cdot
   \mathcal{F}\{U_{\Pi_{l}}^{(i)}\}
\},
\end{equation}
where $U_{\Pi_{l}}^{(i)}$ represents the complex field at depth plane $\Pi_l$ for viewpoint $i$,
and $P^{(i)}$ is the recorded hologram.
As shown in ~\refFig{phase_consistent}, this consistency becomes apparent when examining the reconstructed wavefront across different viewpoints.
In real life, light maintains intrinsic phase offsets at emission that are tied to scene geometry,
while the observed phase relationships at the sensor vary with viewpoint through wave propagation and superposition~\cite{jang2024waveguide}.
Complex-valued holographic radiance field attempts to mimic this physical behavior,
ensuring that the amplitude and phase differences between scene elements remain invariant, leading to physically-consistent interference and diffraction effects.
This represents a paradigm shift—from treating the amplitude and phase as computational variables applied after intensity determination on the hologram plane,
to modeling them as an intrinsic and meaningful property of the 3D scene itself.
By combining the Lagrangian perspective of 3DGS with holography,
we achieve modeling and rendering of a complex-valued holographic radiance field, eliminating the need for expensive per-configuration, per-primitive wave calculations.

\section{Implementation}
We employ the ground-truth images and rendered depth information to generate target reconstruction images using the multiplane generation pipeline by Kavaklı \etal \shortcite{kavakli2023multicolor}.
Additionally, we employ the fused SSIM proposed by \cite{Taming3DGS2024Sig} to speed up the training process.
For densification, we adopt the original strategy from \3DGS, with the exceptions that the densifying frequency is 300 steps and the opacity resetting is replaced by the regularization method proposed by~\citet{rota2024revising}.

We choose Adan~\cite{xie2024adan} as the optimizer ($\beta_{1} = 0.9, \beta_{2} = 0.99$) and
train our complex-valued Gaussians for 20000 steps using parameter-specific learning rates:
$0.005$ for scales, $0.0025$ for both amplitude and phase, $0.025$ for opacity, $0.001$ for rotation,
and a base learning rate of $0.01$ for means and plane assignment probabilities.
Our choice of optimizer follows~\cite{zhang2024gaussianimage}, while the learning rate is adopted from the original \3DGS implementation.
We apply CosineAnnealingLR~\cite{CosineAnnealingLR} only to the learning rates of means and plane assignment probabilities (minimum $0.00001$) for smoother convergence,
while keeping the others fixed throughout training.
For holographic configuration, we follow the most recent literature~\cite{shi2022end, shi2021towards, aksit2023holobeam}
and choose a propagation distance of 2~$mm$, volume depth of 4~$mm$, pixel pitch of 3.74~$\mu m$, and wavelengths of 639, 532, and 473~$nm$.
All experiments are conducted on a NVIDIA A100 80G GPU.
For details on dataset-specific camera viewpoint sampling strategies, please refer to Supplementary~\refSupSec{camera_sampling}.
\begin{table*}[!t]
\centering
\footnotesize
\caption{Quality comparison on NeRF Synthetic and LLFF.
3DGS + U-Net (GT) refers to the \textit{Image Quality Ground Truth} that requires recalculation of hologram per viewpoint.
3DGS + U-Net (Vary) represents the \textit{Viewpoint Relationship Variation} without recalculation.
Metrics are reported as PSNR/SSIM/LPIPS, respectively. Bold indicates best performance excluding GT baseline.
For more information about baselines, please refer to \refSec{baseline_select}.}

\vspace{0.3cm}
\begin{tabular}{l@{\hskip 6pt}|@{\hskip 6pt}c@{\hskip 4pt}c@{\hskip 4pt}c@{\hskip 4pt}c@{\hskip 4pt}c@{\hskip 4pt}c@{\hskip 4pt}c@{\hskip 4pt}c@{\hskip 6pt}|@{\hskip 6pt}c}
\toprule
\multirow{2}{*}{Method} & \multicolumn{9}{c}{NeRF Synthetic (Test Resolution 800×800)} \\
 & chair & drums & ficus & hotdog & lego & materials & mic & ship & Mean \\
\midrule
3DGS + U-Net (GT) & 29.3/0.91/0.13 & 28.9/0.90/0.14 & 31.3/0.95/0.10 & 30.4/0.94/0.15 & 28.1/0.89/0.15 & 29.1/0.92/0.14 & 32.7/0.95/0.10 & 29.1/0.90/0.18 & 29.9/0.92/0.14 \\
3DGS + U-Net (Vary) & 7.2/0.15/0.71 & 9.8/0.12/0.77 & 8.1/0.18/0.72 & 8.5/0.14/0.85 & 7.9/0.09/0.81 & 7.8/0.16/0.75 & 8.4/0.19/0.79 & 9.3/0.11/0.79 & 8.4/0.14/0.77 \\
Our Method & \textbf{27.6}/\textbf{0.89}/\textbf{0.14} & \textbf{26.1}/\textbf{0.86}/\textbf{0.17} & \textbf{28.0}/\textbf{0.91}/\textbf{0.13} & \textbf{29.1}/\textbf{0.90}/\textbf{0.16} & \textbf{24.3}/\textbf{0.83}/\textbf{0.21} & \textbf{28.3}/\textbf{0.88}/\textbf{0.17} & \textbf{29.2}/\textbf{0.91}/\textbf{0.14} & \textbf{24.9}/\textbf{0.81}/\textbf{0.23} & \textbf{27.1}/\textbf{0.87}/\textbf{0.17} \\
\bottomrule
\end{tabular}

\vspace{0.5cm}
\begin{tabular}{l|ccccccc|c}
\toprule
\multirow{2}{*}{Method} & \multicolumn{8}{c}{LLFF (Test Resolution 960×640)} \\
 & fern & flower & fortress & horns & orchids & room & trex & Mean \\
\midrule
3DGS + U-Net (GT) & 29.6/0.90/0.35 & 29.4/0.89/0.32 & 27.4/0.88/0.37 & 28.8/0.91/0.37 & 27.5/0.84/0.39 & 28.9/0.93/0.36 & 26.1/0.83/0.41 & 28.3/0.88/0.37 \\
3DGS + U-Net (Vary) & 8.3/0.17/0.78 & 7.1/0.13/0.82 & 8.8/0.11/0.85 & 9.2/0.19/0.79 & 5.7/0.12/0.83 & 8.9/0.16/0.80 & 7.4/0.12/0.86 & 7.9/0.14/0.82 \\
Our Method & \textbf{27.9}/\textbf{0.78}/\textbf{0.39} & \textbf{28.1}/\textbf{0.80}/\textbf{0.37} & \textbf{25.2}/\textbf{0.73}/\textbf{0.45} & \textbf{26.7}/\textbf{0.79}/\textbf{0.43} & \textbf{25.1}/\textbf{0.73}/\textbf{0.46} & \textbf{27.3}/\textbf{0.81}/\textbf{0.41} & \textbf{24.5}/\textbf{0.75}/\textbf{0.45} & \textbf{26.4}/\textbf{0.77}/\textbf{0.42} \\
\bottomrule
\end{tabular}
\label{tbl:combined_results}
\end{table*}

\section{Evaluation}
We evaluate our method on three standard datasets,
including NeRF Synthetic~\cite{mildenhall2021nerf}, LLFF~\cite{mildenhall2019llff}, and Mip-NeRF 360~\cite{barron2022mip}.

\begin{table}[!htbp]
\centering
\caption{Comparison of inference times across different Gaussian primitive based hologram synthesis methods at 800×800 resolution.
Inference time for 3DGS + U-Net includes per-view recomputation, consisting of \3DGS rendering and network inference for each queried viewpoint.}
\label{tbl:inference_time}
\begin{footnotesize}
\begin{tabular}{lcc}
\toprule
Method & Number of $\mathcal{G}s$ & Inference Time \\
\midrule
\multirow{2}{*}{Our Method} & 200K & 10 ms \\
& 5M & 69 ms \\
\midrule
\multirow{2}{*}{3DGS + U-Net~\cite{chen2025view}} & 200K & 8 ms \\
& 5M & 29 ms \\
\midrule
\multirow{3}{*}{\GWS (Fast) \cite{choi2025gaussian}} & 15K & > 3 s \\
& 200K & > 40 s \\
& 5M & > 15 min \\
\midrule
\multirow{3}{*}{\GWS (Exact) \cite{choi2025gaussian}} & 15K & > 1 min \\
& 200K & > 13 min \\
& 5M & > 5 hrs \\
\bottomrule
\end{tabular}
\end{footnotesize}
\end{table}
\subsection{Baseline Selection}
\label{sec:baseline_select}

We compare our method against a two-stage baseline that generates holograms using a U-Net~\cite{ronneberger2015u} from RGBD images rendered from a pretrained intensity-based 3DGS scene.
The baseline is selected for several reasons. First, although neural network estimates holograms that are not scene-geometric aware,
it achieves millisecond-level inference, substantially faster than optimization-based \CGH methods.
Second, despite introducing unnatural defocus effects and a limited eye box, the network's denoising capability suppresses speckle noise and artifacts commonly found in the other CGH methods,
yielding the highest quantitative image quality.
As such, this method represents a theoretical upper bound on achievable image quality under ideal conditions.
We therefore treat 3DGS + U-Net (GT) as the \textit{Image Quality Ground Truth} for benchmarking.
In addition, we evaluate a \textit{Viewpoint Relationship Variation} setting, denoted 3DGS + U-Net (Vary),
where holograms are generated without recalculation when the relationship between scene geometry and viewpoint changes.

From a computational perspective, 3DGS + U-Net requires rendering RGBD images from the pretrained 3DGS scene at each queried viewpoint,
followed by neural inference to generate the hologram.
Similarly, \GWS first renders the 2DGS scene,
then computes closed-form wave propagation for each Gaussian primitive at the queried viewpoint. Both adopt an Eulerian paradigm,
recalculating holograms for every novel view at camera plane.
We use 3DGS + U-Net as the primary baseline because the source code of \GWS is currently unavailable; our reimplementation yields lower image quality,
and its reconstructions exhibit different defocus blur characteristics than the official implementation, making the comparison inaccurate and unfair.

\subsection{Inference Time}
~\refTbl{inference_time} demonstrates our method's efficiency.
Compared with \GWS~\cite{choi2025gaussian}, our method is inherently more scalable, achieving 30x-10,000x speedup while maintaining view consistency.
While 3DGS + U-Net~\cite{chen2025view} offers faster rendering, it requires recalculation of the hologram per viewpoint, which has no scene geometry-awareness.
Even with a large number of Gaussians (5M), our method maintains reasonable performance (69 $ms$),
achieving a balance between computational efficiency and maintaining scene geometry-aware representations.
\refFig{infer_time} illustrates the scalability of our method; the render time remains relatively constant across
resolutions due to tile-based rasterization, while both memory usage and render time scale linearly with the number of
depth planes and Gaussian primitives, demonstrating the computational efficiency of our method compared to traditional \CGH methods
that rely on per-primitive diffraction calculation.
For ablation study of the computational contribution of tile-based rasterization and FFT-based propagation, please refer to Supplementary~\refSupSec{speedup_ablation}.

\subsection{Complex Field Discontinuities: Beyond Intensity-Based Smooth Interpolation}
\label{sec:sampling_density}

\subsubsection{Quantitative Analysis. }
\refTbl{combined_results} and \ref{tbl:mip360} show the training images performance of our method and 3DGS + U-Net on the NeRF Synthetic,
LLFF, and Mip-NeRF 360 datasets, respectively.
We evaluate the image quality using PSNR, SSIM, and LPIPS~\shortcite{zhang2019LPIPS}.

The \textit{Viewpoint Relationship Variation} results show that 3DGS + U-Net (Vary) fails to generate correct holograms when viewpoint changes without recalculation,
with PSNR dropping by 21.5\,dB, 20.4\,dB, and 19.6\,dB on NeRF Synthetic, LLFF, and Mip-NeRF 360 datasets, respectively, and SSIM decreases by over 0.74 across all 17 scenes.
In contrast, our method maintains robust performance across novel views, achieving 18.7\,dB, 18.5\,dB, and 13.9\,dB PSNR improvements
over 3DGS + U-Net (Vary).

For the \textit{Image Quality Ground Truth}, our method achieves promising performance on the NeRF Synthetic and LLFF datasets,
with PSNR gaps of 2.8\,dB and 1.9\,dB and SSIM differences of 0.05 and 0.11, respectively, compared to 3DGS + U-Net (GT).
However, performance degradation of our method becomes significant on Mip-NeRF 360 dataset with 5.7\,dB PSNR drop and 0.42\,SSIM decrease.
This progressive degradation from controlled synthetic scenes to complex in-the-wild environments stems from the fundamental
challenge of representing spatially discontinuous complex fields to model interference and diffraction.
For more analysis of gaussian distribution and statistics of holograms rendered by our method, please refer to Supplementary~\refSupSec{distribution_analysis}.

\subsubsection{Complex Field Discontinuities. }
\label{sec:sampling_density_explanation}
In intensity-based radiance fields, the goal is to model view-dependent appearance through opacity and color.
This representation handles sparse viewpoint sampling effectively because intensity variations generally follow smooth,
low-frequency patterns that can be interpolated between viewpoints.
However, modeling complex-valued holographic radiance fields introduce a challenge related to motion parallax.
For a camera translation, motion parallax causes objects at different depths to experience different amounts of apparent displacement.
Objects at depths $d_1$ and $d_2$ exhibit displacements of:
\begin{equation}
    \Delta s_1 = \frac{f \cdot \Delta t}{d_1}, \quad \Delta s_2 = \frac{f \cdot \Delta t}{d_2},
\end{equation}
where $f$ is the focal length and $\Delta t$ is the camera translation.
When $d_2 \gg d_1$, the motion parallax difference $\Delta s_1 - \Delta s_2$ becomes significant.
This motion parallax translates directly into phase differences in complex-valued holographic radiance fields.
Between consecutive camera viewpoints, these Gaussian primitives will accumulate different phase shifts:
\begin{equation}
    \Delta\varphi_1 = \frac{2\pi}{\lambda} \Delta s_1, \quad \Delta\varphi_2 = \frac{2\pi}{\lambda} \Delta s_2.
\end{equation}
The issue arises when adjacent pixels correspond to objects with vastly different depths,
creating sharp spatial discontinuities in phase distribution.
Consider two Gaussian primitives at depths $d_1$ and $d_2$
that are spatially adjacent in the hologram plane, the contrasting phase evolution during camera motion will introduce visual artifacts.
For adjacent pixels separated by distance $\Delta x$, the spatial phase gradient is defined as:
\begin{equation}
\frac{\partial \varphi}{\partial x} = \frac{\Delta\varphi_2 - \Delta\varphi_1}{\Delta x} = \frac{2\pi}{\lambda} \frac{f \cdot \Delta t \cdot (d_2 - d_1)}{d_1 d_2 \Delta x}.
\end{equation}
When the scene contains large depth discontinuities ($d_2 \gg d_1$), the spatial phase gradient will become very large.
This creates interpolation challenges absent in intensity-based representations.
Unlike intensity values, which are continuous and monotonic, phase values wrap at $2\pi$ boundaries.
Large spatial phase gradients cause adjacent pixels to differ by multiple cycles,
creating ambiguity in interpolation paths that intensity-based representations do not encounter.
To evaluate our method against complex field discontinuities,
we carefully choose and analyze challenging scenes with significant depth variations from commonly used datasets that exhibit different motion parallax characteristics.

\setlength{\intextsep}{1.5pt}
\setlength{\columnsep}{5pt}
\begin{figure}[t!]
   \centering
   \includegraphics[width=0.47\textwidth]{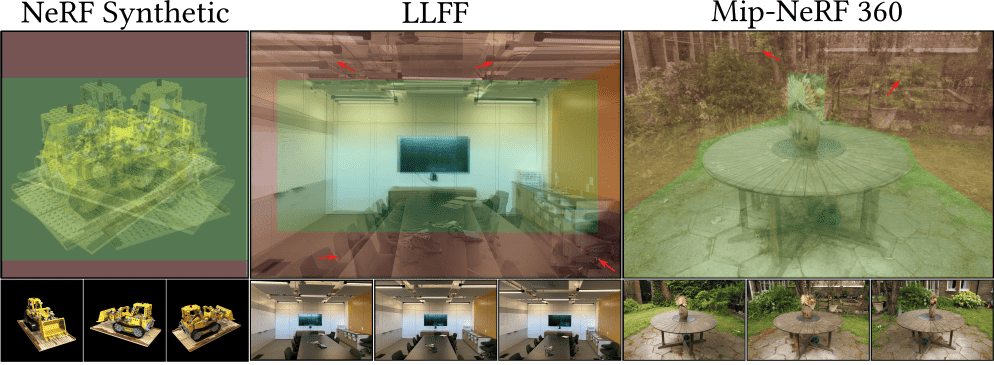}
   \caption{Motion parallax across datasets:
   NeRF Synthetic (left) shows a controlled scene with minimal parallax;
   LLFF (center) shows an indoor scene with vast-moving foreground motion (red region);
   Mip-NeRF 360 (right) shows an in-the-wild scene with wide background shifts (red region).
   Regions exhibiting moderate motion parallax that does not cause phase discontinuities in our experiments are marked in green.
   }
   \label{fig:parallax}
\end{figure}
As shown in \refFig{parallax}, the NeRF Synthetic dataset benefits from controlled capture conditions with relatively uniform depth distributions within each scene.
In contrast, LLFF and Mip-NeRF 360 datasets contain in-the-wild scenes with larger depth variations.
These scenes create sharp phase boundaries where the complex field changes abruptly in between viewpoints,
breaking the spatial coherence assumptions of intensity-based representations.
While the 3DGS + U-Net method maintains visual quality by first modeling geometry and appearance through intensity-based \3DGS
(which handles motion parallax effectively) and then estimating hologram as a secondary task,
this two-stage \CGH approach sacrifices the scene geometry-awareness across different viewpoints.
This limitation highlights the trade-off between the spatial coherence requirements of intensity-based representations
and the phase discontinuities introduced by motion parallax in complex-valued holographic radiance fields.
\begin{table}[ht!]
\centering
\footnotesize
\caption{
Quality comparison on Mip-NeRF 360.
3DGS + U-Net (GT) refers to the \textit{Image Quality Ground Truth} that requires recalculation of hologram per viewpoint.
3DGS + U-Net (Vary) represents the \textit{Viewpoint Relationship Variation} without recalculation.
Metrics are reported as PSNR/SSIM/LPIPS, respectively. }
\begin{tabular}{l|cc}
\toprule
\multirow{2}{*}{Method} & \multicolumn{2}{c}{Mip-NeRF 360 (Test Resolution 960×640)} \\
 & garden & kitchen \\
\midrule
3DGS + U-Net (GT)       & 25.9/0.87/0.35 & 28.1/0.91/0.34 \\
3DGS + U-Net (Vary)     & 7.9/0.11/0.72 & 6.9/0.13/0.78 \\
Our Method              & 20.2/0.45/0.61 & 22.4/0.49/0.53 \\
\bottomrule
\end{tabular}
\label{tbl:mip360}
\end{table}

\subsection{Qualitative Analysis}
\begin{figure}[t!]
   \centering
   \includegraphics[width=0.47\textwidth]{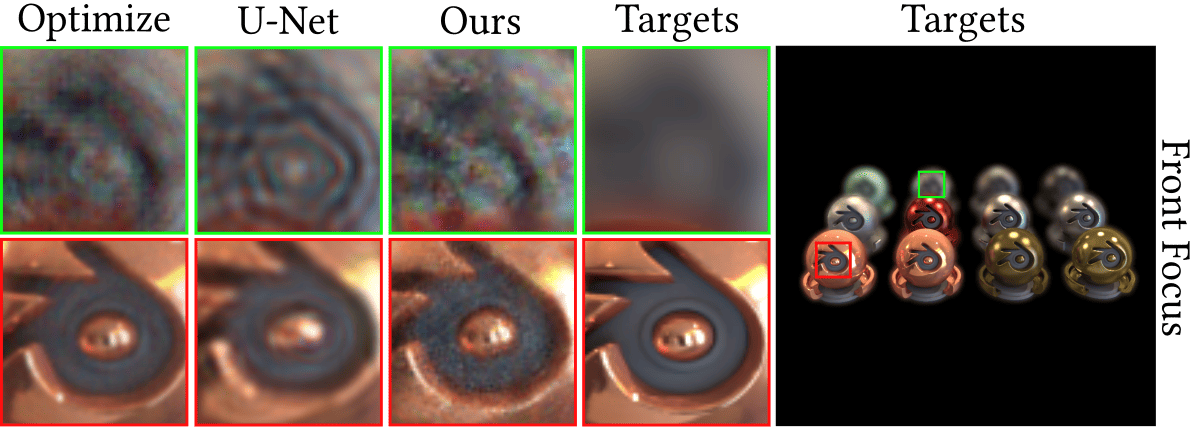}
   \caption{Defocus blur comparison in holographic reconstructions between optimization~\cite{kavakli2023multicolor},
   3DGS + U-Net~\cite{chen2025view}, and our method against target images.}
   \label{fig:defocus}
\end{figure}

\paragraph{Defocus Blur Comparison. } \refFig{defocus} compares simulated defocus blur across three \CGH methods: optimization, 3DGS + UNet, and ours.
Our method generates perceptually more plausible defocus blur, similar to the optimization-based result without per-view recalculation.
3DGS + UNet's result suffers from a structured, fringing effect in defocus blur, which is commonly found in learned \CGH methods~\cite{shi2021towards, chen2025view}.
\setlength{\intextsep}{0.5pt}
\setlength{\columnsep}{10pt}

\begin{figure*}[ht!]
   \centering
   \includegraphics[width=0.99\textwidth]{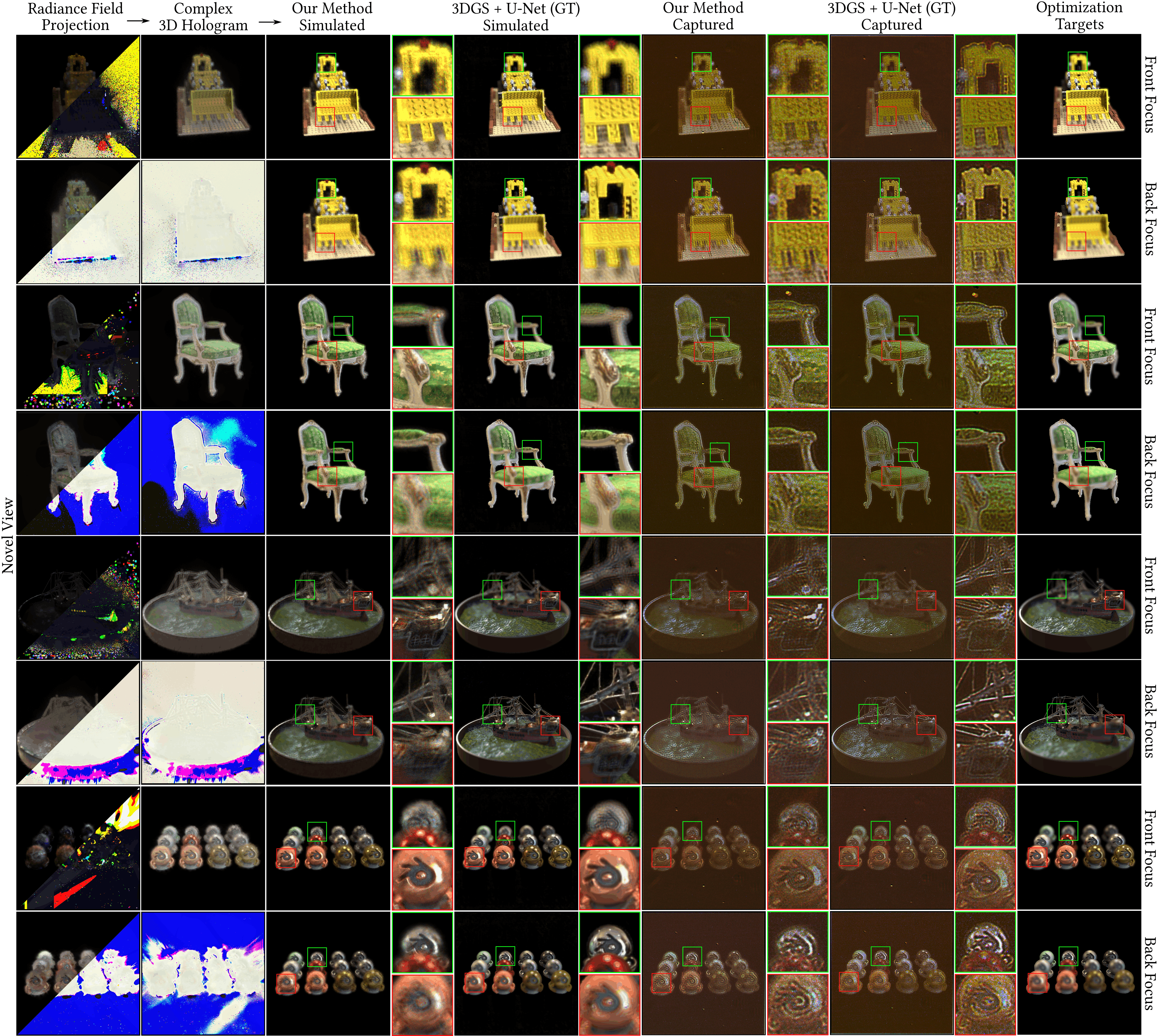}
   \caption{Novel-view comparison between our method and the 3DGS + U-Net baseline on multiple scenes from the NeRF Synthetic dataset (lego, chair, ship, materials).
   Here GT represents the \textit{Image Quality Ground Truth}.
   The first two columns show the radiance field projections and their rendered complex 3D holograms.
   The central columns present a side-by-side evaluation of our method's simulated results and experimentally captured results against the baseline.
   The rightmost column displays the target images used as optimization objectives in our method.}
   \label{fig:result_synthetic}
\end{figure*}

\begin{figure*}[ht!]
   \centering
   \includegraphics[width=0.99\textwidth]{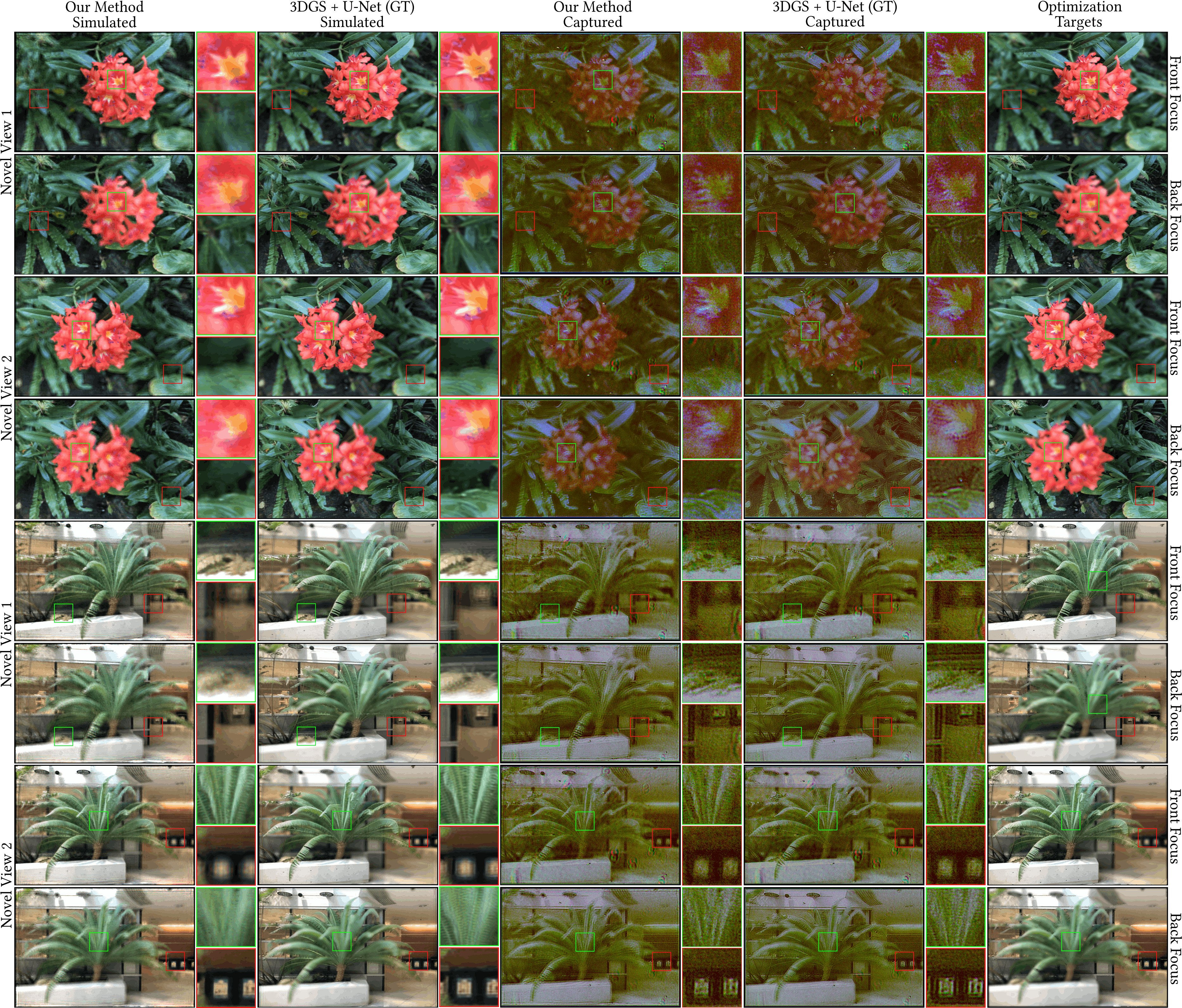}
   \caption{Novel-view comparison between our method and the 3DGS + U-Net baseline on multiple scenes from the LLFF dataset (flower, fern).
   Here GT represents the \textit{Image Quality Ground Truth}.
   The central columns present a side-by-side evaluation of our method's simulated results and experimentally captured results against the baseline.
   The rightmost column displays the target images used as optimization objectives in our method.}
   \label{fig:result_LLFF1}
\end{figure*}

\begin{figure*}[ht!]
   \centering
   \includegraphics[width=0.99\textwidth]{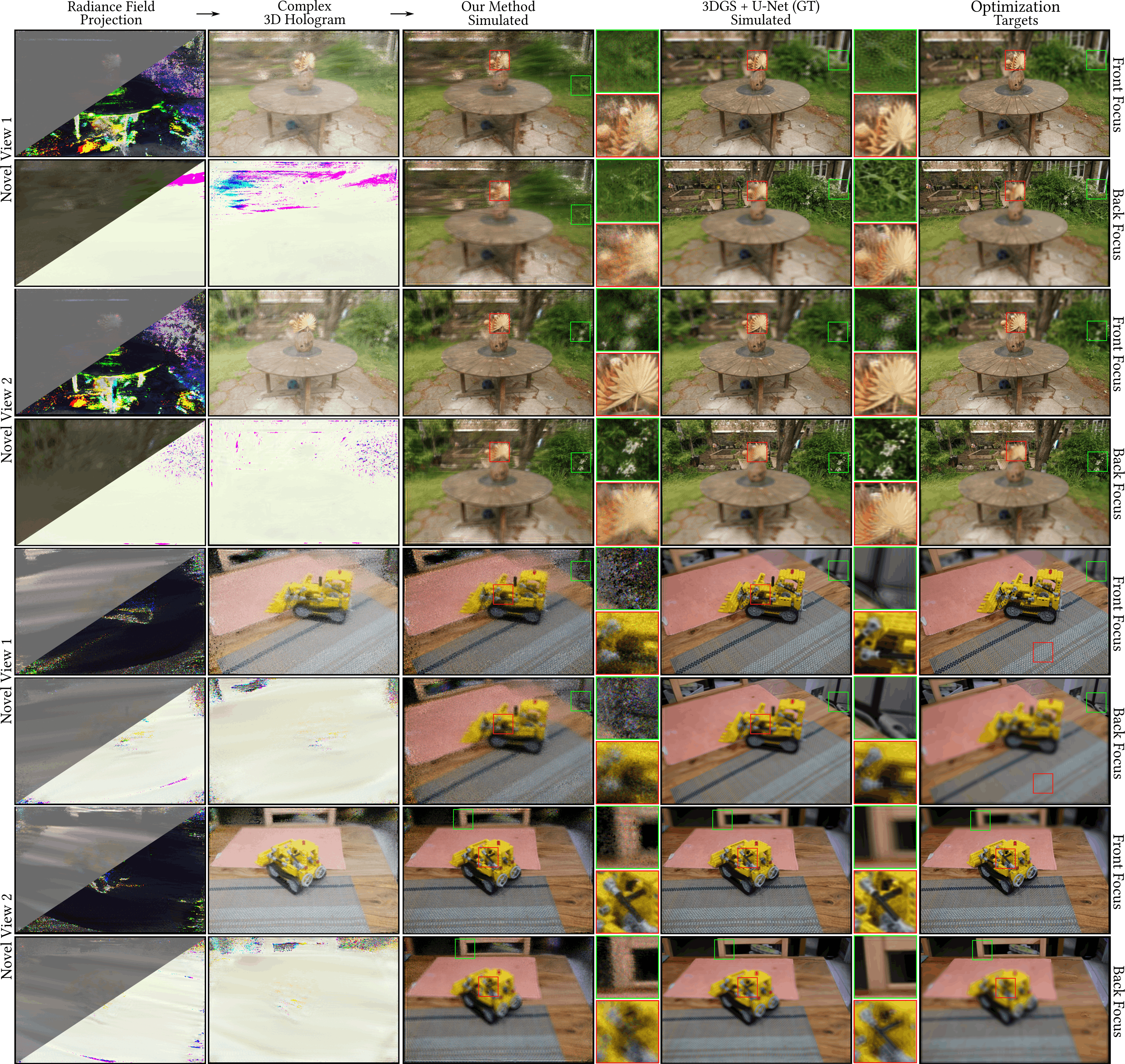}
   \caption{Novel-view comparison of our method on multiple scene from the Mip-NeRF 360 dataset (garden, kitchen).
   The first two columns show the radiance field projections and their rendered complex 3D holograms.
   The central columns present our method and 3DGS + UNet's simulated results.
   The rightmost column displays the target images used as optimization objectives in our method.}
   \label{fig:result_mip360}
\end{figure*}

\paragraph{Novel View Image Quality Analysis. }
We evaluate novel view image quality on the NeRF Synthetic, LLFF, and Mip-NeRF 360 datasets—shown in \refFig{result_synthetic}, \refFig{result_LLFF1}, and \refFig{result_mip360}, respectively—using both simulation and experimental captures.
For the NeRF Synthetic dataset, our method achieves visually consistent results across different datasets and on-par image quality compared to the baseline.

For the LLFF dataset, our method preserves image quality effectively in regions with stable depth content, particularly for background elements.
However, artifacts become apparent in areas containing near-field objects that exhibit rapid motion across viewpoints,
such as foreground furniture and close-range structural elements.
These artifacts manifest as phase inconsistencies and visual distortions,
primarily in regions where objects undergo significant displacement due to motion parallax.

For the Mip-NeRF 360 dataset,
the most significant quality loss of our method occurs in distant background regions that experience substantial motion across large viewpoint changes,
such as far-field vegetation and architectural structures.
Objects at intermediate and near depths maintain reasonable reconstruction quality.
This depth-dependent performance indicates that limitations arise from phase discontinuities caused by motion parallax rather than deficiencies
in our complex-valued representations, as evidenced by the preserved quality in spatially coherent depth regions.
For more simulated and captured results, please refer to Supplementary~\refSupSec{extra_result}.

\paragraph{Phase-only Hologram Conversion.}
To display our complex-valued holograms on a phase-only \SLM,
we leverage a propagation-based optimization approach that converts amplitude-phase into phase-only representations.
Given our complex hologram $P = A e^{j\varphi}$ with amplitude $A$ and phase $\varphi$,
we aim to obtain a corresponding phase-only hologram $P_{\varphi} = e^{j\varphi_{\text{opt}}}$
where the amplitude is unity and the phase $\varphi_{\text{opt}}$ is optimized.
We propagate both holograms to the same set of depth planes and minimize
their propagation differences.
For each depth plane $\Pi_l$, we compute the propagated complex fields
\begin{equation}
\begin{aligned}
U_{l} &= \mathcal{F}^{-1}\{H_{-Z_l}(f_x,f_y) \cdot \mathcal{F}\{P\}\},  \\
U_{\varphi,l} &= \mathcal{F}^{-1}\{H_{-Z_l}(f_x,f_y) \cdot \mathcal{F}\{P_{\varphi}\}\},
\end{aligned}
\end{equation}
and the optimization minimizes the difference between the real and imaginary components across all depth planes
\begin{equation}
\mathcal{L}_{\varphi} = \frac{1}{L} \sum_{l=1}^{L} \|U_{l} - U_{\varphi,l}\|^2 + \mathcal{L}_{SSIM}(U_{l}, U_{\varphi,l}).
\end{equation}
Future variants could adopt learned approaches, such as U-Net and others, to replace optimization and enable more efficient conversion.

\section{Discussion and Future Work}
\subsection{Limitations}
\paragraph{Wave Propagation Overhead. }
Although our method renders 3D holograms effectively,
the rendering process involves splatting Gaussian primitives onto multiple depth planes, followed by wave propagation using the band-limited \ASM{}.
The band-limited \ASM method is not as efficient as differential splatting, particularly at high resolutions,
as it involves multiple high-resolution \FFT{}s to compute accurate wave propagation,
which inevitably increases the computation and the memory cost.
A more efficient light propagation method~\cite{Zhan2025CV2DGaussian} is required to further reduce the memory cost and match the rendering speed of intensity-based \3DGS.

\paragraph{Motion Parallax. }
Our method's reliance on Gaussian primitives for smooth spatial interpolation creates limitations in scenes with significant motion parallax.
Objects at dramatically different depths exhibit varying displacements during viewpoint changes,
generating sharp phase discontinuities that violate the smoothness assumptions of intensity-based representations.
This explains our degraded results on in-the-wild datasets with large depth variations.
Future work could explore hierarchical Gaussians~\cite{wang2025freetimegs} to adaptively model phase discontinuities at multiple scales.

\paragraph{Incoherent Illumination. }
Our method operates under the assumption of fully coherent illumination, where wavefronts exhibit stable phase relationships and produce predictable interference patterns.
In contrast, most natural light sources are only partially coherent or incoherent, with phase correlations that degrade over time or across spatial extents.
Our model's reliance on coherent light leads to sensitivity to depth-induced phase artifacts,
and partial coherence may help mitigate these discontinuities, highlighting an important direction for future work.

\paragraph{Learned Plane Assignment.} The plane assignment probabilities in our method are a computational latent variable rather than a physically-accurate depth labeling.
The consistency across interpolated novel viewpoints is learned implicitly from the training view distribution,
rather than guaranteed by a physical model. While effective in practice, this approach is a pragmatic compromise.
Future work should explore more elegant and physically grounded alternatives.

\subsection{Future Work}
Our work opens the door to potential advances in holographic rendering technology.
\paragraph{Scene Relighting. }
By directly modeling the complex-valued holographic radiance fields, a future variant of our method could support scene relighting~\cite{2019GuoToG}.
Building on prior research in reflectance displays~\cite{2014GlasnerRelight},
our approach could extend spatially varying reflectance functions through learned 3D structure and phase information that responds to illumination changes passively and reactively.

\paragraph{Pupil-aware Holography. }
A future variant could explore estimating complex-valued holographic radiance fields that dynamically adapt to pupil movements under novel views~\cite{chakravarthula2022pupil}.
Moreover, our method inherently contains light field information within the complex-valued 3D representations.
With appropriate training objectives, a future extension of our approach could directly support real-time 4D CGH rendering~\cite{kim2024holographic} without requiring explicit light field extraction.

\paragraph{Holographic Camera. }
Another promising direction lies in bridging synthetic complex-valued holographic radiance fields with real-world holographic acquisition.
Recent advances in incoherent holographic cameras~\cite{yu2023deep,li2025real} can capture both intensity and phase information from real 3D scenes under natural illumination without coherent laser sources.
Integrating such technology with our framework could enable modeling the true wave optics characteristics of real objects, moving beyond current RGB reproduction to authentic physical phase properties.

\paragraph{Physically-informed Color Representation. }
In the physical world, light amplitude is naturally characterized by its spectral distribution, not through mathematical constructs like \SH.
To maintain scene geometry-aware representations, our complex-valued holographic radiance field uses only RGB triplets to define the intrinsic
amplitude of Gaussians to align with the intrinsic phase values. Despite this simplification, our experiments demonstrate that
complex-valued holographic radiance field preserves view-dependent reflectance properties of 3D scenes without requiring high-dimensional view-dependent features.
Compared to \SH, which typically contains 48 parameters, our method requires only 6 (3 for amplitude, 3 for phase),
resulting in reduced model size and memory footprint.
Future work could examine the theoretical basis of this observation and assess whether physically plausible parameterizations further enhance radiance field expressiveness.

\paragraph{Conclusion. }
Our work introduces a novel approach to support holographic representations in Gaussian Splatting by utilizing our novel 3D complex-valued Gaussians as primitives.
We demonstrate that this method leads to a coherent novel view synthesis that is faithful to the geometry of a 3D scene in terms of both amplitude and phase.
These findings from our work demonstrate the potential for more physically-accurate 3D scene representations,
helping holograms to replace 2D images and holographic displays to be compatible and practical as 3D screens in the future.

\begin{acks}
The authors thank Dr. Josef Spjut and Dr. Mike Roberts for providing valuable suggestions in the early phases.
Seung-Hwan acknowledges funding from the National Research Foundation of Korea (NRF) grants funded by the Korea government (MSIT) (RS-2024-00438532, RS-2023-00211658) and the Ministry of Education through the Basic Science Research Program (2022R1A6A1A03052954), as well as grants from the Institute of Information \& Communications Technology Planning \& Evaluation (IITP) funded by the Korea government (MSIT) (No. RS-2024-0045788) and the IITP-ITRC (Information Technology Research Center) program (IITP-2026-RS-2024-00437866).
\end{acks}

\bibliographystyle{ACM-Reference-Format}
\bibliography{references}

\setlength{\intextsep}{0.5pt}
\setlength{\columnsep}{10pt}
%
%

\clearpage
{\centering\Large\bfseries Supplementary Material\par}
\vspace{1em}
\appendix

\section{Hardware Image}
\begin{figure}[ht!]
  \centering
  \includegraphics[width=0.45\textwidth]{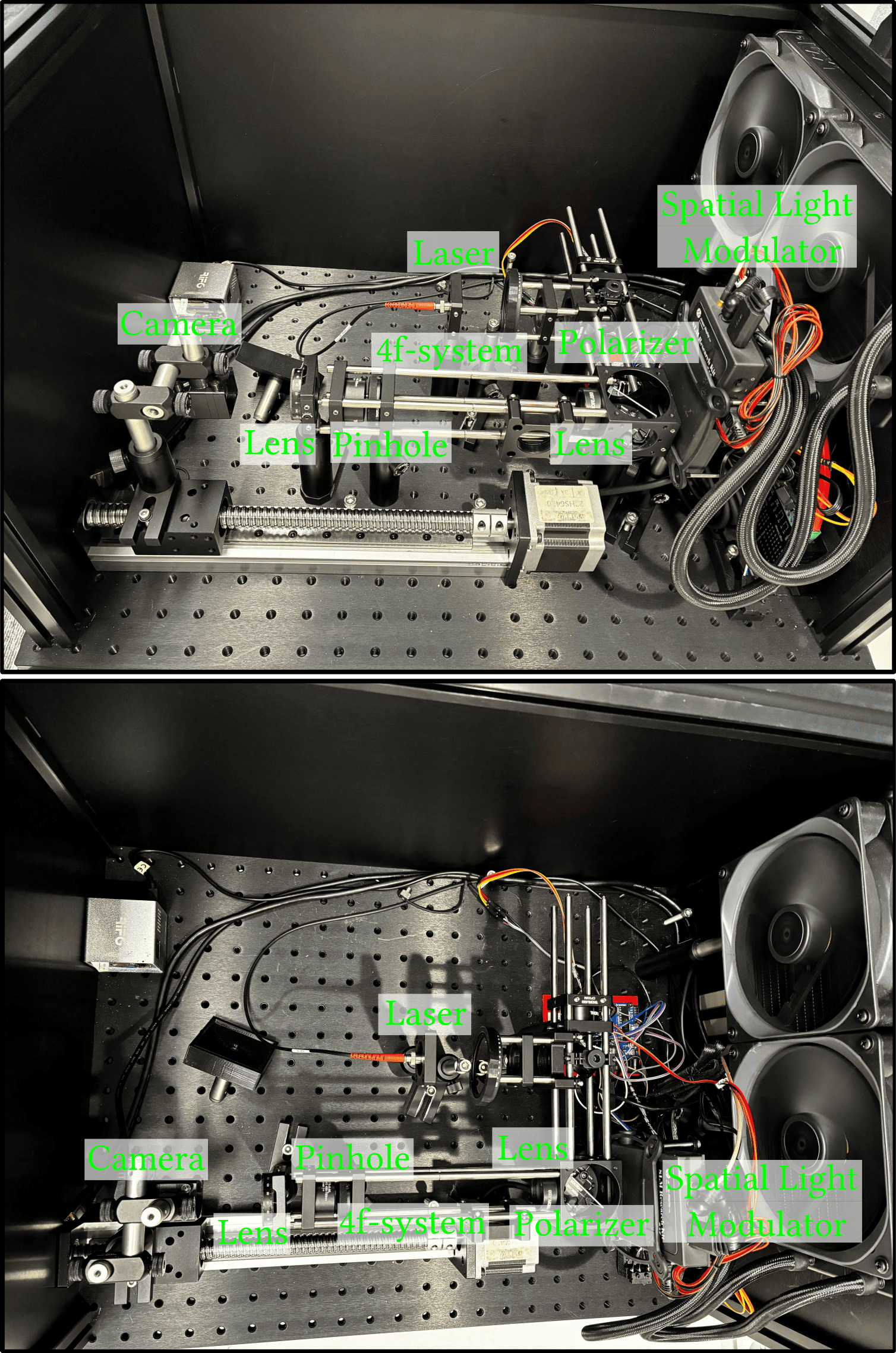}
  \caption{The photograph of the holographic display prototype (Jasper JD7714) used in evaluating holograms generated by our model.}
  \label{fig:hardware_jasper}
\end{figure}

\refFig{hardware_jasper} shows the photograph of the holographic display prototype we used in this paper.
The optical path of our display prototype begins with a laser light source (LASOS MCS4),
which integrates three individual laser lines. The emitted light from a single-mode fibre is collimated using a
Thorlabs LA1708-A plano-convex lens with a 200~mm focal length.
This linearly polarized, collimated beam is then directed by a beamsplitter (Thorlabs BP245B1)
toward our phase-only \SLM, the Jasper JD7714 (2400×4094, 3.74~$\mu$m).
The modulated beam subsequently passes through a lens system comprising Thorlabs LA1908-A and LB1056-A,
with focal lengths of 500~mm and 250~mm, respectively. Following this, a pinhole aperture (Thorlabs SM1D12)
is positioned at the focal plane of the lenses. Finally, we capture the holographic reconstructions using a lensless image sensor
(Point Grey GS3-U3-23S6M-C USB 3.0), which is mounted on an X-stage (Thorlabs PT1/M) with a travel range of 0 to 25~mm and
a positioning precision of 0.01~mm.

\section{Wave Propagation}
In this section, we provide additional insights into the implementation of our proposed complex-valued holographic radiance field.
\subsection{Wave Propagation From a Single 3D Gaussian Primitive}
Each Gaussian $\mathcal{G}_n$ can be considered as a small volume in $\Omega$ that emits a coherent wavefield.
Intuitively, we may approximate the total field from $\mathcal{G}_n$ at hologram pixel $\PPixel$'s coordinate $(x,y)$ by an integral of infinitesimal point contributions at the surface of the Gaussian.
\textbf{To simplify the problem, if we only consider $\mathcal{G}_n$'s center at $\mathbf{x}_n = (X_n, Y_n, Z_n)$},
the distance $D_n$ from the $\mathcal{G}_n$ to the hologram pixel at $(x,y)$ is
\begin{equation}
   D_n = \sqrt{
       (x - X_n)^2 + (y - Y_n)^2 + Z_n^2
   }
   \label{eq:distance_nblob}
\end{equation}
For a point light source emitting spherical waves, the phase at the hologram pixel is
\begin{equation}
   \boldsymbol{\varphi}_{n}^{\text{prop}} = \frac{2\pi}{\lambda}D_n \mod 2\pi
   \label{eq:phase_from_blob}
\end{equation}
Under paraxial approximation, phase can be described using a quadratic function
\begin{equation}
\boldsymbol{\varphi}_{n}^{\text{approx}} = \frac{2\pi}{\lambda}\left(Z_n + \frac{(x - X_n)^2 + (y - Y_n)^2}{2Z_n}\right) \mod 2\pi
\label{eq:phase_paraxial}
\end{equation}
The resulting complex field from Gaussian $\mathcal{G}_n$ at the hologram pixel is
\begin{equation}
U_n = A_n\exp(j\boldsymbol{\varphi}_{n}^{\text{prop}}), \quad A_n = \mathbf{c}_n\,\boldsymbol{\alpha}_n
\label{eq:complex_field_from_blob}
\end{equation}
where $A_n$ includes the Gaussian's $\boldsymbol{\alpha}_n$ weighted amplitude, and $\mathbf{c}_n$ represents the inherent wave amplitude of the Gaussian.
The full complex field integration of $\mathcal{G}_n$ can be derived by expanding this point response to the Gaussian's spatial extent $\mathbf{S}_n$ and rotation $\mathbf{R}_n$ by densely discretizing the surface of the Gaussian.
This yields a volume of wavefront footprint recorded at the hologram plane $P$.
Without considering the occlusion, the total field at the hologram pixel can be defined as:
\begin{equation}
   U^{\text{total}} = \sum_{n=1}^{N} U_n
   \label{eq:total_field_naussians}
\end{equation}
One may average over the sub-samples to compute the field at hologram pixel $\PPixel$, or keeping them explicit for higher accuracy modeling.
\noindent \subsection{Occlusion Aware Field Recording.}
\begin{figure}[ht!]
   \centering
   \includegraphics[width=0.39\textwidth]{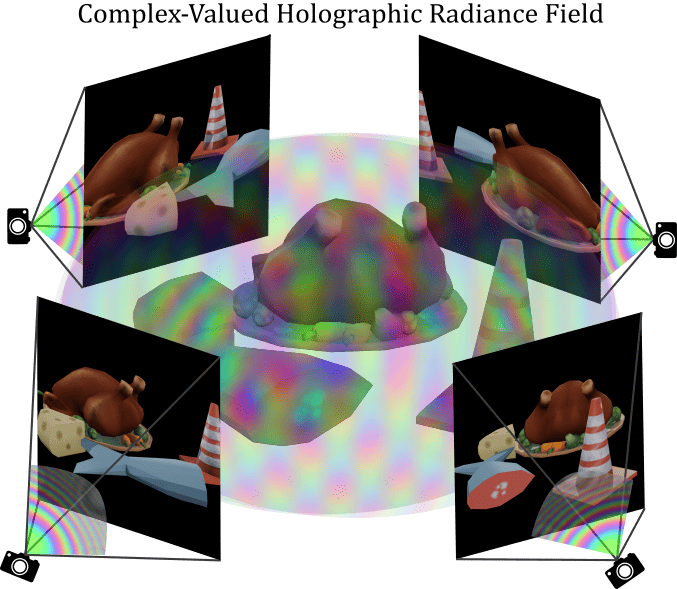}
   \caption{Our method assumes the radiance field is a wave emitter, instead of a ray tracer.}
   \label{fig:wave_emitter}
\end{figure}

In a pure intensity world, to account for fore-ground occlusion, the frontmost Gaussian along the line of sight tends to occlude anything behind it,
and people apply alpha-blending to represent transparent pixels.
For simulating physical wave propagation in a hologram, occlusion is subtler—each "surface" can block or partially block wavefronts behind it.
One approach to accurately model occlusion is to reconstruct a triangle surface mesh by connecting the neighboring points of Gaussians and performing ray casting from each point to the hologram plane $P$.
The wavefronts carried by rays that hit the surface mesh are excluded from the hologram calculation.
A binary visibility mask can be generated from this process to mimic foreground-occlusion at the hologram plane.
However, both point-based wave simulation and ray-triangle detection of the entire scene are computationally expensive,
making them a less favorable choice in real-time inference.
As illustrated in \refFig{wave_emitter}, in our implementation, we use alpha-blending with transmittance to efficiently approximate occlusion effects,
each 3D Gaussian $\mathcal{G}_n$ can be considered as a small volume in $\Omega$ that emits a coherent wavefield, rather than the more computationally expensive ray-casting approach described above.

\section{Training Camera Viewpoint Sampling}
\label{supplementary:camera_sampling}

During training, camera viewpoints are sampled differently depending on the dataset characteristics to ensure the learned scene representation supports
consistent rendering across viewpoints within each dataset's coverage.

For the NeRF Synthetic dataset, we randomly sample camera poses facing towards the object within the Blender coordinate system,
maintaining the same radial distance distribution as the original dataset while randomizing the azimuth and elevation angles.
This generates a continuous distribution of training viewpoints around the object.

For the LLFF dataset, we directly use the existing camera viewpoints provided in the dataset,
as the captured views already provide sufficient coverage of the indoor scenes with manageable motion parallax.

For the Mip-NeRF 360 dataset, we employ a two-stage approach due to complex field discontinuities challenges arising from extreme motion parallax in the original wide-baseline captures.
First, we train an intensity-based 3DGS scene using the provided dataset images.
We uniformly sample 480 camera views covering a full 360° rotation, corresponding to 0.75° angular spacing for training, and evaluate novel-view rendering on the intermediate unseen viewpoints between sampled angles.
We then use this trained 3DGS scene as a reference to randomly sample camera viewpoints facing towards the scene center, similar to the NeRF Synthetic sampling strategy.
As a way round method, this approach generates training views with reduced inter-view motion parallax in the near scene compared to the original Mip-NeRF 360 captures,
mitigating the severe complex field discontinuities that would otherwise cause substantial image quality degradation, but it does not help with the far field.

\section{Differentiable Complex-Valued Gaussian Rendering}
\label{supplementary:cuda_gradient}
In this section, we provide the pseudo code and a detailed exposition of both the forward computation and the backward gradient derivation for our Differentiable Complex-Valued Rasterizer.

\subsection{Notation}
\begin{itemize}
\item $\mathbf{x}_n \in \mathbb{R}^3$ - 3D Gaussian center positions in world space
\item $\mathbf{x}_n^{\text{proj}} \in \mathbb{R}^2$ - Projected 2D Gaussian centers in screen space
\item $\mu_x, \mu_y$ - Components of the 2D mean position
\item $\Sigma \in \mathbb{R}^{3 \times 3}$ - 3D covariance matrix in world space
\item $\Sigma' \in \mathbb{R}^{2 \times 2}$ - 2D covariance matrix in screen space
\item $\Sigma'^{-1}_{ij}$ - Elements of the inverse 2D covariance matrix (where $i,j \in \{0,1\}$)
\item $d_x = x - \mu_x, d_y = y - \mu_y$ - Distance from pixel to Gaussian center
\item $\boldsymbol{\rho}_{n,l}$ - Plane assignment probability of Gaussian $n$ for plane $l$ (controls contribution to each rendering plane)
\item $\boldsymbol{\varphi}_{n}$ - Phase value for Gaussian $n$ (controls complex field phase)
\item $\boldsymbol{\alpha}_n$ - Opacity value for Gaussian $n$
\item $\boldsymbol{\alpha}_n^{\text{base}}$ - Base opacity value before applying the Gaussian function
\item $T_n$ - Accumulated transmittance at Gaussian $n$ (product of previous transparency values)
\item $z_{\text{norm}}$ - Normalized depth value between 0 and 1
\end{itemize}

\subsection{Optimization and Densification Algorithm}
Our optimization and densification algorithms extend the standard \3DGS~\cite{kerbl20233d} approach to handle complex-valued parameters and multi-plane assignments. The key modifications include phase parameter optimization, plane probability learning with \STE, and complex-valued loss computation as summarized in Algorithm~\ref{alg:complex_optimization}.

\begin{algorithm}[ht!]
\caption{Optimization and Densification for Complex-Valued 3DGS}
\label{alg:complex_optimization}
\footnotesize
\begin{algorithmic}[1]
\Require $w, h$: width and height of the training images
\Require $L$: number of depth planes for multi-plane rendering
\State $M \leftarrow$ SfM Points \Comment{Positions}
\State $\mathbf{S}, \mathbf{C}, \boldsymbol{\alpha} \leftarrow$ InitAttributes() \Comment{Scales, Amplitudes, Opacities}
\State $\boldsymbol{\varphi} \leftarrow$ InitPhase() \Comment{Phase Parameters}
\State $\boldsymbol{\rho}' \leftarrow$ InitPlaneProbs($L$) \Comment{Plane Assignment Logits}
\State $i \leftarrow 0$ \Comment{Iteration Count}
\While{not converged}
    \State $V, I \leftarrow$ SampleTrainingView() \Comment{Camera $V$ and Image}
    \State $\boldsymbol{\rho} \leftarrow$ STE($\boldsymbol{\rho}'$) \Comment{Hard Assignment via STE}
    \State $U_{\text{complex}} \leftarrow$ ComplexRasterize($M, \mathbf{S}, \mathbf{C}, \boldsymbol{\alpha}, \boldsymbol{\varphi}, \boldsymbol{\rho}, V$) \Comment{Alg. \ref{alg:complex_rasterizer}}
    \State $P \leftarrow$ PropagateToHologram($U_{\text{complex}}$) \Comment{Multi-plane ASM}
    \State $I_{\text{recon}} \leftarrow$ BackPropagate($P$) \Comment{Reconstruct Images}
    \State $\mathcal{L} \leftarrow$ ComplexLoss($I, I_{\text{recon}}$) \Comment{$\mathcal{L}_{\text{recon}} + \mathcal{L}_{\text{SSIM}}$}
    \State $M, \mathbf{S}, \mathbf{C}, \boldsymbol{\alpha}, \boldsymbol{\varphi}, \boldsymbol{\rho}' \leftarrow$ Adan($\nabla\mathcal{L}$) \Comment{Backprop \& Step}
    \If{IsRecentIteration()}
        \ForAll{Gaussians $(\mu, \Sigma, c, \alpha, \varphi, \rho')$ \textbf{in} $(M, \mathbf{S}, \mathbf{C}, \boldsymbol{\alpha}, \boldsymbol{\varphi}, \boldsymbol{\rho}')$}
            \If{$\alpha < \epsilon$ \textbf{or} IsTooLarge($\mu, \Sigma, c, \alpha$)} \Comment{Pruning}
                \State RemoveGaussian()
            \EndIf
            \If{$\nabla_{\mu}\mathcal{L} > \tau_p$} \Comment{Densification}
                \If{$\|\mathbf{S}\| > \tau_s$} \Comment{Over-reconstruction}
                    \State SplitGaussian($\mu, \Sigma, c, \alpha, \varphi, \rho'$)
                \Else \Comment{Under-reconstruction}
                    \State CloneGaussian($\mu, \Sigma, c, \alpha, \varphi, \rho'$)
                \EndIf
            \EndIf
        \EndFor
    \EndIf
    \State $i \leftarrow i + 1$
\EndWhile
\end{algorithmic}
\end{algorithm}

The key extensions to the standard algorithm include: (1) Phase parameter $\boldsymbol{\varphi}$ initialization with random values in $[0, 2\pi)$; (2) Plane assignment probabilities $\boldsymbol{\rho}'$ initialized uniformly and converted to hard assignments via \STE during forward pass; (3) Complex-valued rasterization producing multi-plane complex fields; (4) Hologram generation through multi-plane \ASM propagation; (5) Phase-aware densification where both split and clone operations preserve phase relationships by adding small random perturbations to maintain coherent interference patterns.

\subsection{Details of the Rasterizer}
Our tile-based rasterizer extends the standard 3DGS approach to handle complex-valued rendering across multiple depth planes. The algorithm processes Gaussians in back-to-front order within each tile to ensure correct alpha blending for wave-optical effects, as detailed in Algorithm~\ref{alg:complex_rasterizer}.

\begin{algorithm}[ht!]
\caption{Complex-Valued Tile-Based Rasterization}
\label{alg:complex_rasterizer}
\footnotesize
\begin{algorithmic}[1]
\Require $w, h$: width and height of the image to rasterize
\Require $M, \mathbf{S}$: Gaussian means and covariances in world space
\Require $\mathbf{C}, \boldsymbol{\alpha}, \boldsymbol{\varphi}, \boldsymbol{\rho}$: Amplitudes, opacities, phases, and plane assignments
\Require $V$: view configuration of current camera
\Require $L$: number of depth planes

\Function{ComplexRasterize}{$w, h, M, \mathbf{S}, \mathbf{C}, \boldsymbol{\alpha}, \boldsymbol{\varphi}, \boldsymbol{\rho}, V$}
    \State CullGaussians($M, V$) \Comment{Frustum Culling}
    \State $M', \mathbf{S}' \leftarrow$ ScreenspaceGaussians($M, \mathbf{S}, V$) \Comment{Transform}
    \State $T \leftarrow$ CreateTiles($w, h$) \Comment{$16 \times 16$ Tiles}
    \State $\mathcal{I}, \mathcal{K} \leftarrow$ DuplicateWithDepthKeys($M', T$) \Comment{Back-to-Front Keys}
    \State $\mathcal{K}_s, \mathcal{I}_s \leftarrow$ SortByKeys($\mathcal{I}, \mathcal{K}$) \Comment{Depth Sort}
    \State $\mathcal{R} \leftarrow$ IdentifyTileRanges($T, \mathcal{K}_s$) \Comment{Tile Ranges}
    \State $U_{\text{real}}, U_{\text{imag}} \leftarrow$ InitComplexCanvas($L, w, h$) \Comment{Multi-plane Output}
    \ForAll{Tiles $t$ \textbf{in} $T$}
        \ForAll{Planes $l$ \textbf{in} $L$}
            \ForAll{Pixels $pix$ \textbf{in} $t$}
                \State $T_{\text{acc}} \leftarrow 1.0$ \Comment{Transmittance}
                \State $\text{real}_{\text{acc}}, \text{imag}_{\text{acc}} \leftarrow 0.0, 0.0$ \Comment{Complex Accumulation}
                \State range $\leftarrow$ GetTileRange($\mathcal{R}, t$)
                \For{$g$ \textbf{in} range} \Comment{Back-to-Front Order}
                    \State $w_{\text{plane}} \leftarrow \boldsymbol{\rho}_{g,l}$ \Comment{Plane Assignment}
                    \If{$w_{\text{plane}} < \epsilon_{\text{plane}}$}
                        \State \textbf{continue}
                    \EndIf
                    \State $\alpha_{\text{eff}} \leftarrow \boldsymbol{\alpha}_g \cdot \exp(-\text{power}_{g,\text{pix}}) \cdot w_{\text{plane}}$ \Comment{Effective Alpha}
                    \State $\text{scale} \leftarrow \mathbf{C}_g \cdot \alpha_{\text{eff}} \cdot T_{\text{acc}}$ \Comment{Complex Scale}
                    \State $\text{real}_{\text{acc}} \leftarrow \text{real}_{\text{acc}} + \text{scale} \cdot \cos(\boldsymbol{\varphi}_g)$ \Comment{Real Component}
                    \State $\text{imag}_{\text{acc}} \leftarrow \text{imag}_{\text{acc}} + \text{scale} \cdot \sin(\boldsymbol{\varphi}_g)$ \Comment{Imaginary Component}
                    \State $T_{\text{acc}} \leftarrow T_{\text{acc}} \cdot (1 - \alpha_{\text{eff}})$ \Comment{Update Transmittance}
                    \If{$T_{\text{acc}} < \epsilon_T$}
                        \State \textbf{break} \Comment{Early Termination}
                    \EndIf
                \EndFor
                \State $U_{\text{real}}[l, \text{pix}] \leftarrow \text{real}_{\text{acc}}$
                \State $U_{\text{imag}}[l, \text{pix}] \leftarrow \text{imag}_{\text{acc}}$
            \EndFor
        \EndFor
    \EndFor
    \State \Return $U_{\text{real}} + j \cdot U_{\text{imag}}$
\EndFunction
\end{algorithmic}
\end{algorithm}

The rasterizer incorporates several key innovations:
(1) Back-to-front depth sorting using inverted depth keys to ensure proper alpha blending for wave propagation;
(2) Multi-plane rendering where each Gaussian contributes to planes based on learned assignment probabilities $\boldsymbol{\rho}_{g,l}$;
(3) Complex field accumulation using amplitude-weighted trigonometric functions $\mathbf{C}_g \cdot \cos(\boldsymbol{\varphi}_g)$ and $\mathbf{C}_g \cdot \sin(\boldsymbol{\varphi}_g)$ for real and imaginary components;
(4) Plane-specific early termination when transmittance falls below threshold $\epsilon_T$; (5) Efficient shared memory usage for tile-based processing with complex-valued intermediate storage.

\subsection{Forward Pass}

\subsubsection{Complex Field Generation}
For each pixel $(x,y)$ in each plane $\Pi_l$, the complex field $U$ is:

\begin{equation}
  U_{\Pi_l}(x,y) = \text{real}_{\Pi_l}(x,y) + i \cdot \text{imag}_{\Pi_l}(x,y)
\end{equation}

Where for each Gaussian $\mathcal{G}_n$:
\begin{equation}
  \begin{split}
    \text{real}_{\Pi_l}(x,y) = \sum_n \mathbf{c}_{n} \cdot \boldsymbol{\alpha}_n \cdot T_n \cdot \cos(\boldsymbol{\varphi}_{n}) \\
    \text{imag}_{\Pi_l}(x,y) = \sum_n \mathbf{c}_{n} \cdot \boldsymbol{\alpha}_n \cdot T_n \cdot \sin(\boldsymbol{\varphi}_{n})
  \end{split}
\end{equation}
\subsubsection{Transmittance Calculation}
\begin{equation}
T_n = \prod_{j=1}^{n-1} (1 - \boldsymbol{\alpha}_j)
\end{equation}
\subsubsection{Alpha Calculation}
\begin{equation}
  \boldsymbol{\alpha}_n = \boldsymbol{\alpha}_n^{\text{base}} \cdot \mathcal{G}_n^{\text{proj}} \cdot \boldsymbol{\rho}_{n,l}
\end{equation}
\subsubsection{Gaussian Projection (Exponent Term)}
\begin{equation}
  \begin{split}
    \mathcal{G}_n^{\text{proj}} = \exp\left(-0.5 \cdot (d_x \cdot (\Sigma'^{-1}_{00} \cdot d_x + \Sigma'^{-1}_{10} \cdot d_y) \right. \\
    \left. + d_y \cdot (\Sigma'^{-1}_{01} \cdot d_x + \Sigma'^{-1}_{11} \cdot d_y))\right)
  \end{split}
\end{equation}
Where $d_x = x - \mu_x$ and $d_y = y - \mu_y$.

\subsection{Backward Pass}

\subsubsection{Gradient Flow Overview}

For each parameter $\theta$, the gradient follows the chain rule from the final loss $\mathcal{L}$ to the parameter:
\begin{equation}
\frac{\partial \mathcal{L}}{\partial \theta} = \frac{\partial \mathcal{L}}{\partial P} \cdot \frac{\partial P}{\partial U_{\Pi_l}} \cdot \frac{\partial U_{\Pi_l}}{\partial \theta}
\end{equation}
where $\mathcal{L}$ is the loss and $P$ is the hologram output.

\subsubsection{Detailed Gradient Derivation for Each Parameter}
\begin{enumerate}
\item \textbf{Gradient for rotation parameters (related to $\mathbf{R}_n$)}

The gradient with respect to rotation parameters follows the chain rule:
\begin{align}
\frac{\partial \mathcal{L}}{\partial \mathbf{R}_n} &= \frac{\partial \mathcal{L}}{\partial P} \cdot \frac{\partial P}{\partial U_{\Pi_l}} \cdot \frac{\partial U_{\Pi_l}}{\partial \boldsymbol{\alpha}_n} \cdot \frac{\partial \boldsymbol{\alpha}_n}{\partial \mathcal{G}_n^{\text{proj}}} \cdot \frac{\partial \mathcal{G}_n^{\text{proj}}}{\partial \text{term}} \nonumber \\
&\cdot \frac{\partial \text{term}}{\partial \Sigma'^{-1}} \cdot \frac{\partial \Sigma'^{-1}}{\partial \Sigma'} \cdot \frac{\partial \Sigma'}{\partial \Sigma} \cdot \frac{\partial \Sigma}{\partial \mathbf{R}_n} \cdot \frac{\partial \mathbf{R}_n}{\partial \mathbf{q}}
\end{align}
Key components include:
\begin{enumerate}
\item $\frac{\partial \mathcal{L}}{\partial P}$: Gradient of loss with respect to hologram output, provided by optimization algorithm.

\item $\frac{\partial P}{\partial U_{\Pi_l}}$: Derivative of hologram with respect to complex field, computed via band-limited propagation.

\item $\frac{\partial U_{\Pi_l}}{\partial \boldsymbol{\alpha}_i} = \mathbf{c}_i \cdot T_i \cdot e^{j\cdot\boldsymbol{\varphi}_i} - \sum_{k=i+1}^{N} \mathbf{c}_k \cdot \boldsymbol{\alpha}_k \cdot T_k \cdot \frac{\partial T_k}{\partial \boldsymbol{\alpha}_i} \cdot e^{j\cdot\boldsymbol{\varphi}_k}$ where $\frac{\partial T_k}{\partial \boldsymbol{\alpha}_i} = -T_k/(1-\boldsymbol{\alpha}_i)$ for $i < k$.

\item $\frac{\partial \boldsymbol{\alpha}_n}{\partial \mathcal{G}_n^{\text{proj}}} = \boldsymbol{\alpha}_n^{\text{base}} \cdot \boldsymbol{\rho}_{n,l}$, from $\boldsymbol{\alpha}_n = \boldsymbol{\alpha}_n^{\text{base}} \cdot \mathcal{G}_n^{\text{proj}} \cdot \boldsymbol{\rho}_{n,l}$.

\item $\frac{\partial \mathcal{G}_n^{\text{proj}}}{\partial \text{term}} = -0.5 \cdot \mathcal{G}_n^{\text{proj}}$, from $\mathcal{G}_n^{\text{proj}} = \exp(-0.5 \cdot \text{term})$.

\item $\frac{\partial \text{term}}{\partial \Sigma'^{-1}} = \Delta \cdot \Delta^T$ where $\Delta = (p-\mathbf{x}_n^{\text{proj}})$ is the pixel-mean difference.

\item $\frac{\partial \Sigma'^{-1}}{\partial \Sigma'} = -\Sigma'^{-1} \otimes \Sigma'^{-1}$, following matrix inverse derivative.

\item $\frac{\partial \Sigma'_{ij}}{\partial \Sigma_{kl}} = \sum_{m,n} J_{im} \mathbf{R}_{mk} \mathbf{R}_{nl} J_{jn}$, where $J$ is the projection Jacobian and $\mathbf{R}$ the rotation matrix.

\item $\frac{\partial \Sigma}{\partial \mathbf{R}_{ij}} = \frac{\partial (\mathbf{R}\,\mathbf{S}\,\mathbf{S}^{\top}\,\mathbf{R}^{\top})}{\partial \mathbf{R}_{ij}} = E_{ij} \cdot \mathbf{S} \cdot \mathbf{S}^T \cdot \mathbf{R}^T + \mathbf{R} \cdot \mathbf{S} \cdot \mathbf{S}^T \cdot E_{ji}^T$, where $E_{ij}$ is a standard basis matrix.

\item $\frac{\partial \mathbf{R}_n}{\partial \mathbf{q}} = \frac{\partial \mathbf{R}}{\partial \mathbf{q}}$, where $\mathbf{q}$ is the quaternion representation of rotation.
\end{enumerate}

\item \textbf{Gradient for $\mathbf{x}_n$ (Gaussian centers)}

The gradient for $\mathbf{x}_n$ passes through multiple paths in the computation graph:
\begin{align}
\frac{\partial \mathcal{L}}{\partial \mathbf{x}_n} &= \frac{\partial \mathcal{L}}{\partial P} \cdot \frac{\partial P}{\partial U_{\Pi_l}} \cdot \frac{\partial U_{\Pi_l}}{\partial \boldsymbol{\alpha}_n} \cdot \frac{\partial \boldsymbol{\alpha}_n}{\partial \mathcal{G}_n^{\text{proj}}} \cdot \frac{\partial \mathcal{G}_n^{\text{proj}}}{\partial \text{term}} \nonumber \\
&\cdot \frac{\partial \text{term}}{\partial \text{diff}} \cdot \frac{\partial \text{diff}}{\partial \mathbf{x}_n^{\text{proj}}} \cdot \frac{\partial \mathbf{x}_n^{\text{proj}}}{\partial \mathbf{x}_n^{\text{cam}}} \cdot \frac{\partial \mathbf{x}_n^{\text{cam}}}{\partial \mathbf{x}_n}
\end{align}
Key components include:
\begin{enumerate}
\item $\frac{\partial U_{\Pi_l}}{\partial \boldsymbol{\alpha}_n}$: As described above.
\item $\frac{\partial \boldsymbol{\alpha}_n}{\partial \mathcal{G}_n^{\text{proj}}} = \boldsymbol{\alpha}_n^{\text{base}} \cdot \boldsymbol{\rho}_{n,l}$
\item $\frac{\partial \mathcal{G}_n^{\text{proj}}}{\partial \text{term}} = -0.5 \cdot \mathcal{G}_n^{\text{proj}}$
\item $\frac{\partial \text{term}}{\partial \text{diff}} = \Sigma'^{-1} \cdot \text{diff} + \text{diff} \cdot \Sigma'^{-1}$, where $\text{diff} = (p-\mathbf{x}_n^{\text{proj}})$ is the pixel-mean difference
\item $\frac{\partial \text{diff}}{\partial \mathbf{x}_n^{\text{proj}}} = -I$ (negative identity matrix)
\item $\frac{\partial \mathbf{x}_n^{\text{proj}}}{\partial \mathbf{x}_n^{\text{cam}}}$: Projection Jacobian
\item $\frac{\partial \mathbf{x}_n^{\text{cam}}}{\partial \mathbf{x}_n} = \mathbf{R}$ (camera rotation matrix)
\end{enumerate}
Critically, for pixels $(x,y)$ and Gaussian centers $(\mu_x, \mu_y)$, we set $d_x = x - \mu_x$ and $d_y = y - \mu_y$. The gradient components are:
\begin{equation}
\begin{aligned}
\frac{\partial \text{term}}{\partial \mu_x} &= -(-d_x \cdot \text{inv}_{00} - d_y \cdot \text{inv}_{10}) \\
\frac{\partial \text{term}}{\partial \mu_y} &= -(-d_x \cdot \text{inv}_{01} - d_y \cdot \text{inv}_{11})
\end{aligned}
\end{equation}
Note the negative sign in the implementation that results from $\frac{\partial (x-\mu_x)}{\partial \mu_x} = -1$. This is correctly captured in our final gradient computation.

\item \textbf{Gradient for covariance matrix elements}

The covariance gradient calculation is performed using the chain rule and the matrix inverse derivative formula:
\begin{equation}
\frac{\partial \mathcal{L}}{\partial \Sigma} = -\Sigma^{-1} \cdot \frac{\partial \mathcal{L}}{\partial \Sigma^{-1}} \cdot \Sigma^{-1}
\end{equation}
For the 2D covariance matrix elements, we compute:
\begin{equation}
\begin{aligned}
\frac{\partial \mathcal{L}}{\partial \Sigma_{00}} &= -(bt_{00} \cdot tmp_{00} + bt_{01} \cdot tmp_{10}) \\
\frac{\partial \mathcal{L}}{\partial \Sigma_{01}} &= -(bt_{00} \cdot tmp_{01} + bt_{01} \cdot tmp_{11}) \\
\frac{\partial \mathcal{L}}{\partial \Sigma_{10}} &= -(bt_{10} \cdot tmp_{00} + bt_{11} \cdot tmp_{10}) \\
\frac{\partial \mathcal{L}}{\partial \Sigma_{11}} &= -(bt_{10} \cdot tmp_{01} + bt_{11} \cdot tmp_{11})
\end{aligned}
\end{equation}
where:
\begin{itemize}
\item $bt$ is the transpose of the inverse covariance matrix
\item $tmp$ represents intermediate gradient values from the quadratic form
\end{itemize}
This ensures that the negative sign from the matrix inverse derivative is properly incorporated.

\item \textbf{Gradient for $\mathbf{S}_n$ (Gaussian scales)}

\begin{align}
\frac{\partial \mathcal{L}}{\partial \mathbf{S}_n} &= \frac{\partial \mathcal{L}}{\partial P} \cdot \frac{\partial P}{\partial U_{\Pi_l}} \cdot \frac{\partial U_{\Pi_l}}{\partial \boldsymbol{\alpha}_n} \cdot \frac{\partial \boldsymbol{\alpha}_n}{\partial \mathcal{G}_n^{\text{proj}}} \cdot \frac{\partial \mathcal{G}_n^{\text{proj}}}{\partial \text{term}} \nonumber \\
&\cdot \frac{\partial \text{term}}{\partial \Sigma'^{-1}} \cdot \frac{\partial \Sigma'^{-1}}{\partial \Sigma'} \cdot \frac{\partial \Sigma'}{\partial \Sigma} \cdot \frac{\partial \Sigma}{\partial \mathbf{S}_n}
\end{align}
\begin{enumerate}
\item Components from $\frac{\partial \mathcal{L}}{\partial P}$ through $\frac{\partial \Sigma'}{\partial \Sigma}$ as previously described.
\item $\frac{\partial \Sigma_{ij}}{\partial \mathbf{S}_{n,k}} = 2 \cdot \mathbf{S}_{n,k} \cdot \mathbf{R}_{i,k} \cdot \mathbf{R}_{j,k}$ for anisotropic case.
\end{enumerate}

\item \textbf{Gradient for $\mathbf{c}_n$ (Gaussian amplitudes)}

\begin{equation}
\frac{\partial \mathcal{L}}{\partial \mathbf{c}_n} = \frac{\partial \mathcal{L}}{\partial P} \cdot \frac{\partial P}{\partial U_{\Pi_l}} \cdot \frac{\partial U_{\Pi_l}}{\partial \mathbf{c}_n}
\end{equation}
\begin{enumerate}
\item $\frac{\partial U_{\Pi_l}}{\partial \mathbf{c}_n} = \boldsymbol{\alpha}_n \cdot T_n \cdot e^{j\cdot\boldsymbol{\varphi}_n}$, directly contributing to the complex field.
\end{enumerate}
In the implementation, this becomes:
\begin{equation}
\frac{\partial \mathcal{L}}{\partial \mathbf{c}_n} = \alpha_n \cdot T_{\text{before}} \cdot (\cos(\varphi_n) \cdot \frac{\partial \mathcal{L}}{\partial \text{real}} + \sin(\varphi_n) \cdot \frac{\partial \mathcal{L}}{\partial \text{imag}})
\end{equation}

\item \textbf{Gradient for $\boldsymbol{\varphi}_n$ (Gaussian phases)}

\begin{equation}
\frac{\partial \mathcal{L}}{\partial \boldsymbol{\varphi}_n} = \frac{\partial \mathcal{L}}{\partial P} \cdot \frac{\partial P}{\partial U_{\Pi_l}} \cdot \frac{\partial U_{\Pi_l}}{\partial \boldsymbol{\varphi}_n}
\end{equation}
\begin{enumerate}
\item $\frac{\partial U_{\Pi_l}}{\partial \boldsymbol{\varphi}_n} = j \cdot \mathbf{c}_n \cdot \boldsymbol{\alpha}_n \cdot T_n \cdot e^{j\cdot\boldsymbol{\varphi}_n}$, from the derivative of the complex exponential.
\end{enumerate}
This expands to:
\begin{equation}
\frac{\partial \mathcal{L}}{\partial \boldsymbol{\varphi}_n} = \alpha_n \cdot T_{\text{before}} \cdot \mathbf{c}_n \cdot (-\sin(\varphi_n) \cdot \frac{\partial \mathcal{L}}{\partial \text{real}} + \cos(\varphi_n) \cdot \frac{\partial \mathcal{L}}{\partial \text{imag}})
\end{equation}
Note the negative sign on the sine term, which comes from the derivative of $\cos(\varphi_n)$.

\item \textbf{Gradient for $\boldsymbol{\alpha}_n^{\text{base}}$ (Gaussian base opacities)}

\begin{equation}
\frac{\partial \mathcal{L}}{\partial \boldsymbol{\alpha}_n^{\text{base}}} = \frac{\partial \mathcal{L}}{\partial P} \cdot \frac{\partial P}{\partial U_{\Pi_l}} \cdot \frac{\partial U_{\Pi_l}}{\partial \boldsymbol{\alpha}_n} \cdot \frac{\partial \boldsymbol{\alpha}_n}{\partial \boldsymbol{\alpha}_n^{\text{base}}}
\end{equation}
\begin{enumerate}
\item $\frac{\partial U_{\Pi_l}}{\partial \boldsymbol{\alpha}_n}$: As previously described.
\item $\frac{\partial \boldsymbol{\alpha}_n}{\partial \boldsymbol{\alpha}_n^{\text{base}}} = \mathcal{G}_n^{\text{proj}} \cdot \boldsymbol{\rho}_{n,l}$ from $\boldsymbol{\alpha}_n = \boldsymbol{\alpha}_n^{\text{base}} \cdot \mathcal{G}_n^{\text{proj}} \cdot \boldsymbol{\rho}_{n,l}$.
\end{enumerate}
In the implementation, this is calculated as:
\begin{equation}
\frac{\partial \mathcal{L}}{\partial \boldsymbol{\alpha}_n^{\text{base}}} = \exp(\text{power}) \cdot \boldsymbol{\rho}_{n,l} \cdot \frac{\partial \mathcal{L}}{\partial \boldsymbol{\alpha}_n}
\end{equation}

\item \textbf{Gradient for $\boldsymbol{\rho}_{n,l}$ (plane assignment probabilities)}

\begin{align}
\frac{\partial \mathcal{L}}{\partial \boldsymbol{\rho}_{n,l}} &= \frac{\partial \mathcal{L}}{\partial P} \cdot \frac{\partial P}{\partial U_{\Pi_l}} \cdot \frac{\partial U_{\Pi_l}}{\partial \boldsymbol{\alpha}_n} \cdot \frac{\partial \boldsymbol{\alpha}_n}{\partial \boldsymbol{\rho}_{n,l}}
\end{align}
\begin{enumerate}
\item $\frac{\partial U_{\Pi_l}}{\partial \boldsymbol{\alpha}_n}$: As previously described.
\item $\frac{\partial \boldsymbol{\alpha}_n}{\partial \boldsymbol{\rho}_{n,l}} = \boldsymbol{\alpha}_n^{\text{base}} \cdot \mathcal{G}_n^{\text{proj}}$
\end{enumerate}
In the implementation, the gradient for plane probabilities is:
\begin{equation}
\frac{\partial \mathcal{L}}{\partial \boldsymbol{\rho}_{n,l}} = \boldsymbol{\alpha}_n^{\text{base}} \cdot \exp(\text{power}) \cdot \frac{\partial \mathcal{L}}{\partial \boldsymbol{\alpha}_n}
\end{equation}
\end{enumerate}

\subsubsection{Multi-Plane Transmittance Calculation}
For multiple planes, the transmittance is calculated separately for each plane $l$:
\begin{equation}
T_n^{(l)} = \prod_{j=1}^{n-1}(1-\boldsymbol{\alpha}_j^{(l)}), \quad \text{where} \quad \boldsymbol{\alpha}_j^{(l)} = \boldsymbol{\alpha}_j^{\text{base}} \cdot \mathcal{G}_j^{\text{proj}} \cdot \boldsymbol{\rho}_{j,l}
\end{equation}
The gradient of transmittance for plane $l$ is:
\begin{equation}
\frac{\partial T_k^{(l)}}{\partial \boldsymbol{\alpha}_n^{(l)}} =
\begin{cases}
-T_k^{(l)}/(1-\boldsymbol{\alpha}_n^{(l)}) & \text{if } n < k \\
0 & \text{otherwise}
\end{cases}
\end{equation}
This approach ensures each plane's rendering is treated independently, with plane weights determining which planes a Gaussian contributes to.
The CUDA implementation computes gradients per-pixel and per-Gaussian for each plane, with gradients accumulated atomically across all planes.

\section{Ablation Study of CUDA operations}
\label{supplementary:speedup_ablation}
\begin{figure*}[t]
  \centering
  \includegraphics[width=0.99\textwidth]{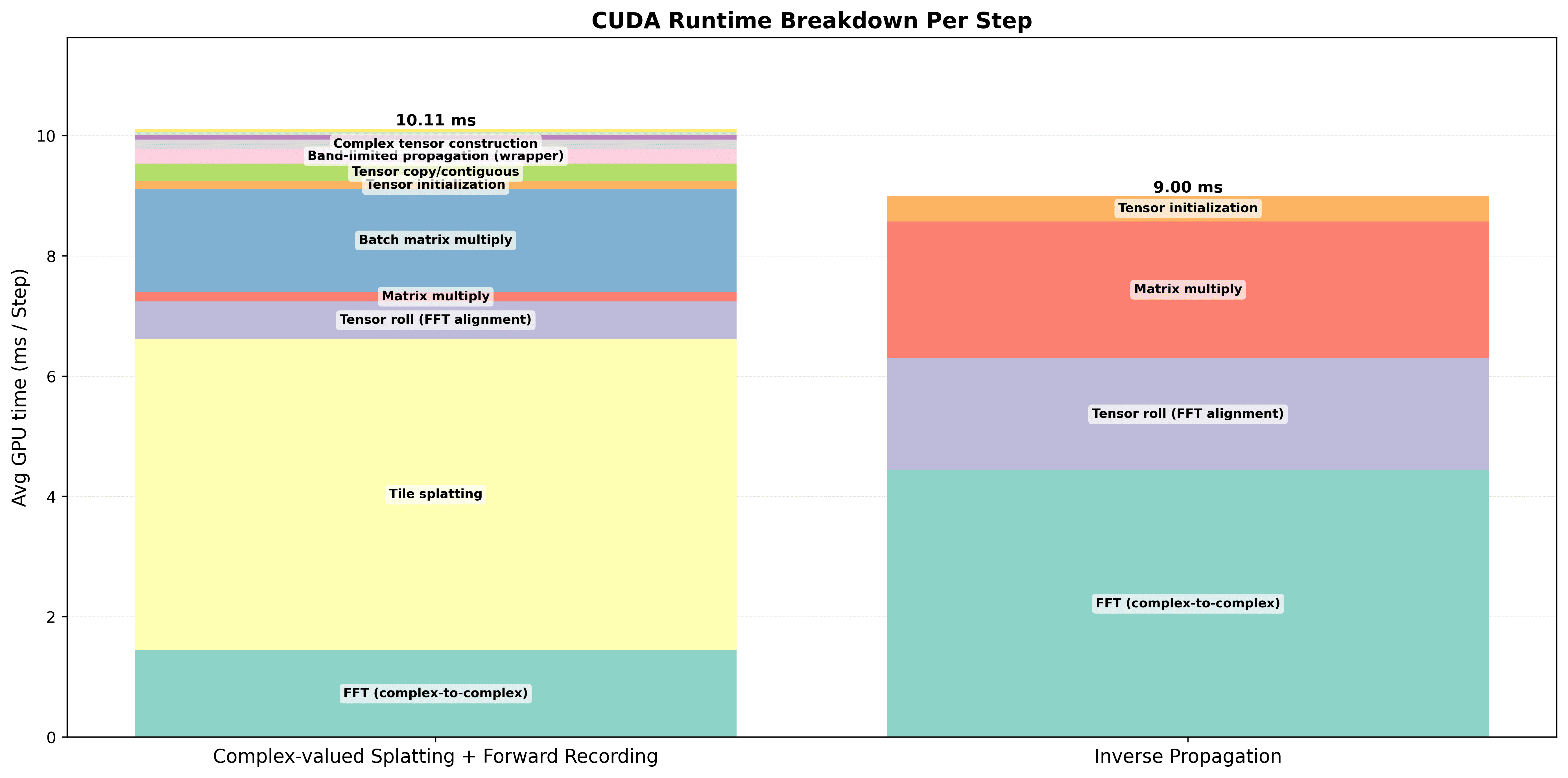}
  \caption{Ablation study of CUDA runtime breakdown per step.}
  \label{fig:gpu_profile_breakdown}
\end{figure*}

This section conduct ablation study on the contribution of our two key method components: tile-based complex-valued rasterization and FFT-based layer propagation.
We measure performance using \textit{torch.profiler}, reporting GPU time.
Each iteration is wrapped by explicit profiling scopes, and the two components are isolated as timing windows;
all CUDA kernels executed within each window are collected via hierarchical traversal and accumulated using self-device time to avoid double-counting.
Per-kernel GPU times are averaged over 50 iterations to obtain a stable per-step runtime breakdown, without inserting any explicit CUDA synchronization that would perturb normal execution.

\refFig{gpu_profile_breakdown} presents a detailed profiling analysis of the computational breakdown for each rendering step.
Averaged by 50 steps, the complex-valued splatting and forward recording step completes in 10.11~ms, with tile-based splatting (SplatTileCuda) accounting for 5.18~ms (51.2\%),
making it the dominant operation. Additional costs include batch matrix operations (1.71~ms, 16.9\%) and FFT operations for band-limited propagation (1.44~ms, 14.3\%).

The inverse propagation step completes in 9.00~ms, dominated by FFT-based complex-to-complex transformations (4.43~ms, 49.2\%)
for multi-plane propagation. Matrix multiplications contribute 2.27~ms (25.3\%), while tensor roll operations for FFT alignment require 1.87~ms (20.7\%).
Together, tile-based rasterization and FFT-based propagation constitute 9.61~ms of the total 19.11~ms pipeline (50.3\%),
demonstrating that these two techniques are the primary computational contributors enabling real-time performance.
The remaining overhead from tensor operations, sorting, and memory management represents 9.50~ms (49.7\%).

\section{Gaussian Distribution And Phase Characteristics Analysis}
\label{supplementary:distribution_analysis}

This section presents a quantitative analysis of the learned complex-valued holographic radiance field, focusing on phase behavior, spatial frequency content,
and their relationship to Gaussian primitive properties.

\subsection{Phase and Amplitude Distributions}

\begin{figure*}[ht!]
   \centering
   \includegraphics[width=0.99\textwidth]{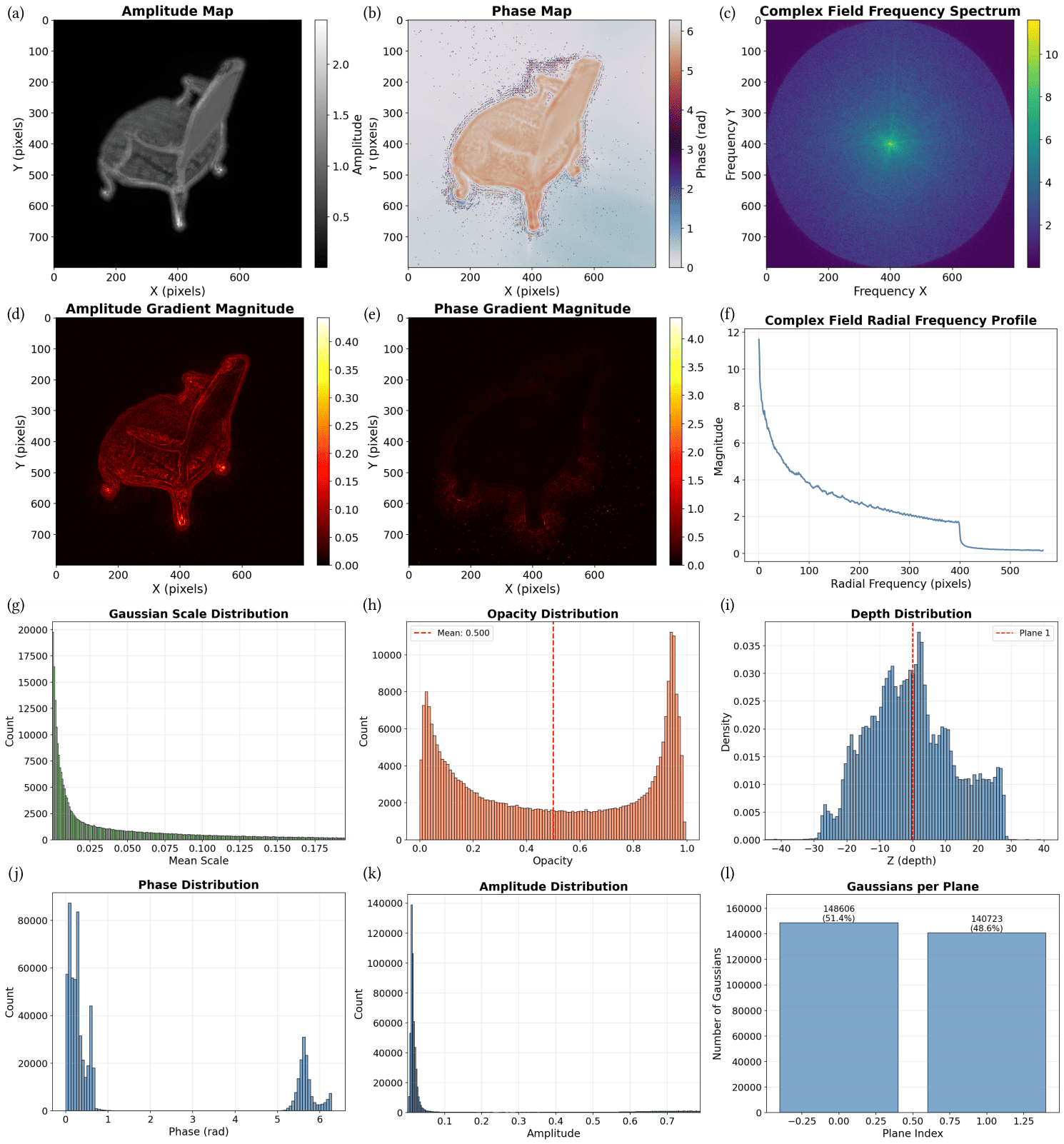}
   \caption{Statistical analysis of learned complex-valued holographic radiance field properties.
            \textbf{(a)} Amplitude map.
            \textbf{(b)} Phase map.
            \textbf{(c)} Fourier-domain energy distribution.
            \textbf{(d)} Amplitude gradient magnitude.
            \textbf{(e)} Phase gradient magnitude.
            \textbf{(f)} Spatial frequency profile.
            \textbf{(g)} Gaussian scale distribution.
            \textbf{(h)} Opacity distribution.
            \textbf{(i)} Depth distribution.
            \textbf{(j)} Phase distribution.
            \textbf{(k)} Amplitude distribution.
            \textbf{(l)} Gaussian assignment across two depth planes.}
   \label{fig:statistics_analysis}
\end{figure*}

The learned hologram phase exhibits a structured distribution that lies between globally smooth and fully random patterns, with a clear bias toward spatial smoothness.
As shown in \refFig{statistics_analysis}(j), the phase histogram is bimodal with peaks near $0$ and $2\pi$, indicating non-uniform but organized phase statistics.

This behavior arises from the coherent superposition of multiple complex-valued Gaussian primitives. Each Gaussian acts as a spatially localized,
low-pass basis function, enforcing continuous phase variation within its finite support. Consequently, individual primitives cannot generate high-frequency, pixel-wise phase fluctuations.

The final phase field is formed by summing many such locally smooth contributions, each associated with an independently learned phase parameter.
Their interference introduces structured, higher-frequency variations while preserving local continuity, yielding phase patterns that are irregular yet not fully random.

Two factors jointly govern this behavior: (1) the Gaussian primitives impose local smoothness due to their finite spatial support; (2) phase parameters $\boldsymbol{\varphi}_n$
are learned per primitive rather than per pixel, inducing spatially correlated phase variations. As illustrated in \refFig{statistics_analysis}(b),
the resulting phase map exhibits extended coherent regions with sharper transitions primarily at overlap regions and object boundaries.

The amplitude distribution \refFig{statistics_analysis}(k) is strongly right-skewed, with most values near zero and a long tail at higher amplitudes.
This reflects the sparsity of the representation, where only a subset of Gaussians contribute significantly at each spatial location. Low amplitudes correspond to background regions,
while the tail captures areas of constructive interference. The amplitude map \refFig{statistics_analysis}(a) confirms that energy is concentrated in object regions.

\subsection{Spatial Coherence and Gradient Analysis}

The low-pass nature of Gaussian primitives induces spatially smooth phase structure, as shown by the phase gradient magnitude in \refFig{statistics_analysis}(e).
Gradients remain low across most of the hologram plane, increasing primarily at object boundaries and depth discontinuities. This behavior follows directly from the continuous Gaussian projections.

Interference between overlapping Gaussians introduces localized high-frequency variations. The amplitude gradient map (\refFig{statistics_analysis}(d))
shows sharp transitions concentrated at object edges, where multiple primitives with differing phases overlap.

\subsection{Spatial Frequency Characteristics}

Fourier-domain analysis \refFig{statistics_analysis}(c,f) reveals a spectrum dominated by low-frequency components, consistent with smooth spatial variations from Gaussian primitives.
Nevertheless, substantial mid-to-high frequency energy is present near the center, encoding fine interference details required for accurate holographic reconstruction.
The spatial frequency profile decays monotonically from DC.

\subsection{Gaussian Primitive Characteristics}

The learned Gaussian scale distribution \refFig{statistics_analysis}(g) is heavily concentrated in the range 0.00--0.05,
with exponential decay toward larger scales. This reflects the densification strategy's preference for small primitives in geometrically complex regions,
enabling high-resolution modeling while maintaining efficiency, this is consistent with traditional 3DGS. The dominance of small-scale Gaussians further promotes local phase coherence due to their limited spatial support.

The opacity distribution \refFig{statistics_analysis}(h) is bimodal, with peaks near 0.0 and 1.0, indicating specialization toward either transparent or opaque states.
The mean opacity near 0.5 reflects a balance between coverage and transmittance, enabling effective modeling of wavefront occlusion via accumulated transmittance.

\subsection{Depth Distribution and Plane Assignment}
The depth distribution \refFig{statistics_analysis}(i) shows Gaussians clustering around predefined depth planes, confirming effective learning of plane assignment probabilities.
This validates the effectiveness of the \STE-based plane assignment.

For multi-plane settings, the plane assignment statistics (\refFig{statistics_analysis}(l)) show balanced utilization across planes (e.g., 51.4\% vs. 48.6\%),
indicating no collapse to a single plane.  As shown in the first row`s zoomed insets of \refFig{gaussian_analysis},
the high-frequency details of Gaussians for front and back focus are gathered at separate depth planes,
demonstrating the effectiveness of the plane splitting of our method.

\subsection{Relationship to Holographic Reconstruction}
\begin{figure*}[ht!]
   \centering
   \includegraphics[width=0.99\textwidth]{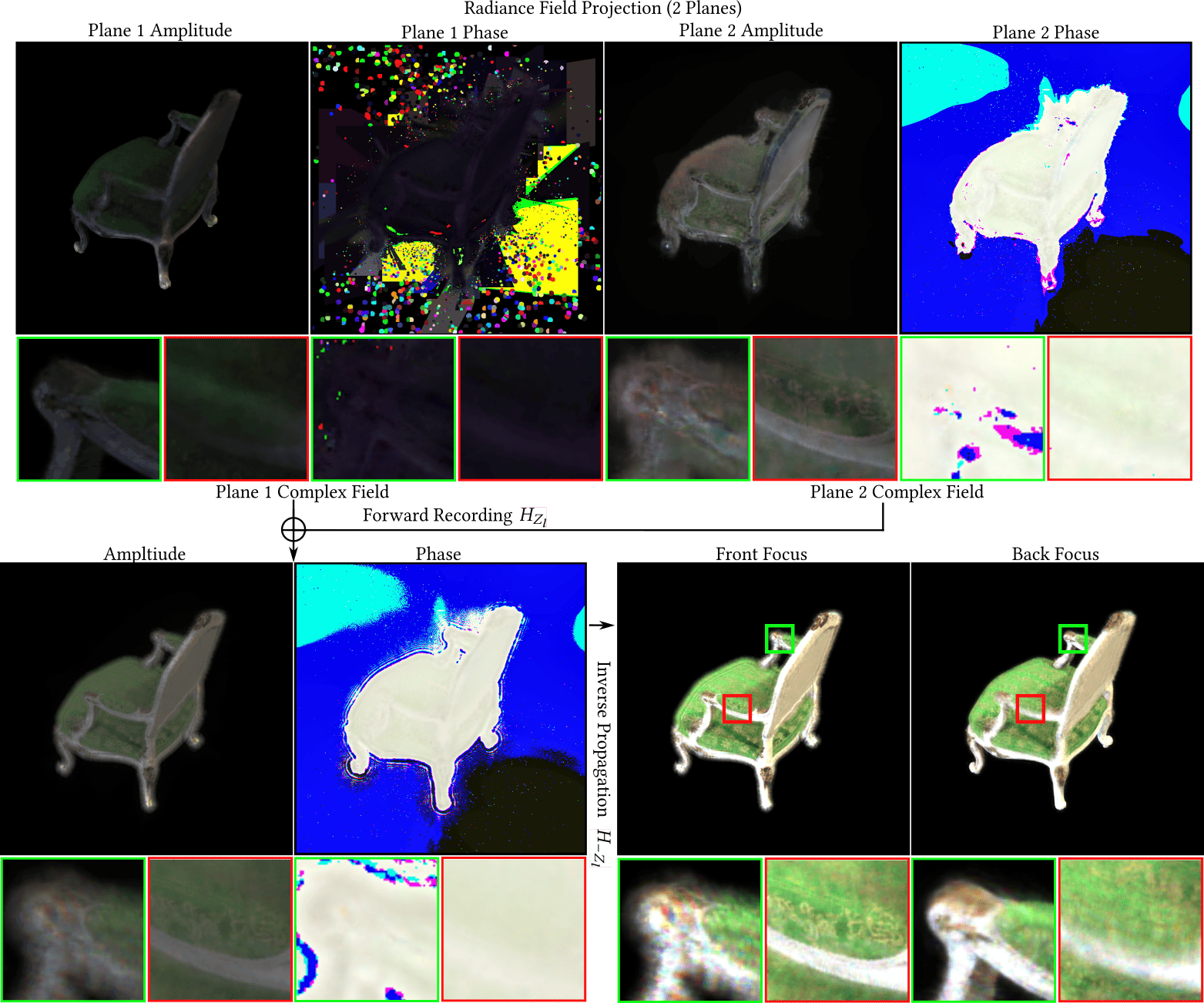}
   \caption{Visualization of complex-valued holographic radiance field rendering pipeline and holographic reconstruction.
   \textbf{Top:} Multi-plane radiance field projections showing amplitude and phase maps for 2 Depth Planes.
   \textbf{Bottom Left} Forward recording process combines plane-specific complex fields via band-limited \ASM ($H_{Z_l}$) to generate the final hologram.
   \textbf{Bottom Right:} Reconstructed Intensities for front and back focus via Inverse Propagation ($H_{-Z_l}$). }
   \label{fig:gaussian_analysis}
\end{figure*}

\refFig{gaussian_analysis} demonstrates how the learned parameters in our method will affect reconstruction. Multi-plane amplitude and phase maps
exhibit coherent phase structure within objects and localized discontinuities at depth boundaries, consistent with the statistical analysis.

During forward recording, plane-specific complex fields are aggregated via band-limited \ASM propagation ($H_{Z_l}$).
Inverse propagation ($H_{-Z_l}$) produces reconstructed intensities with natural defocus blur.
In-focus regions remain sharp, while out-of-focus areas exhibit smooth diffraction blur without structured artifacts existed in the existing learned CGH methods.

\subsection{Novel View Synthesis Progression}

\begin{landscape}
\begin{figure}[p]
    \centering
    \includegraphics[width=1.2\textwidth]{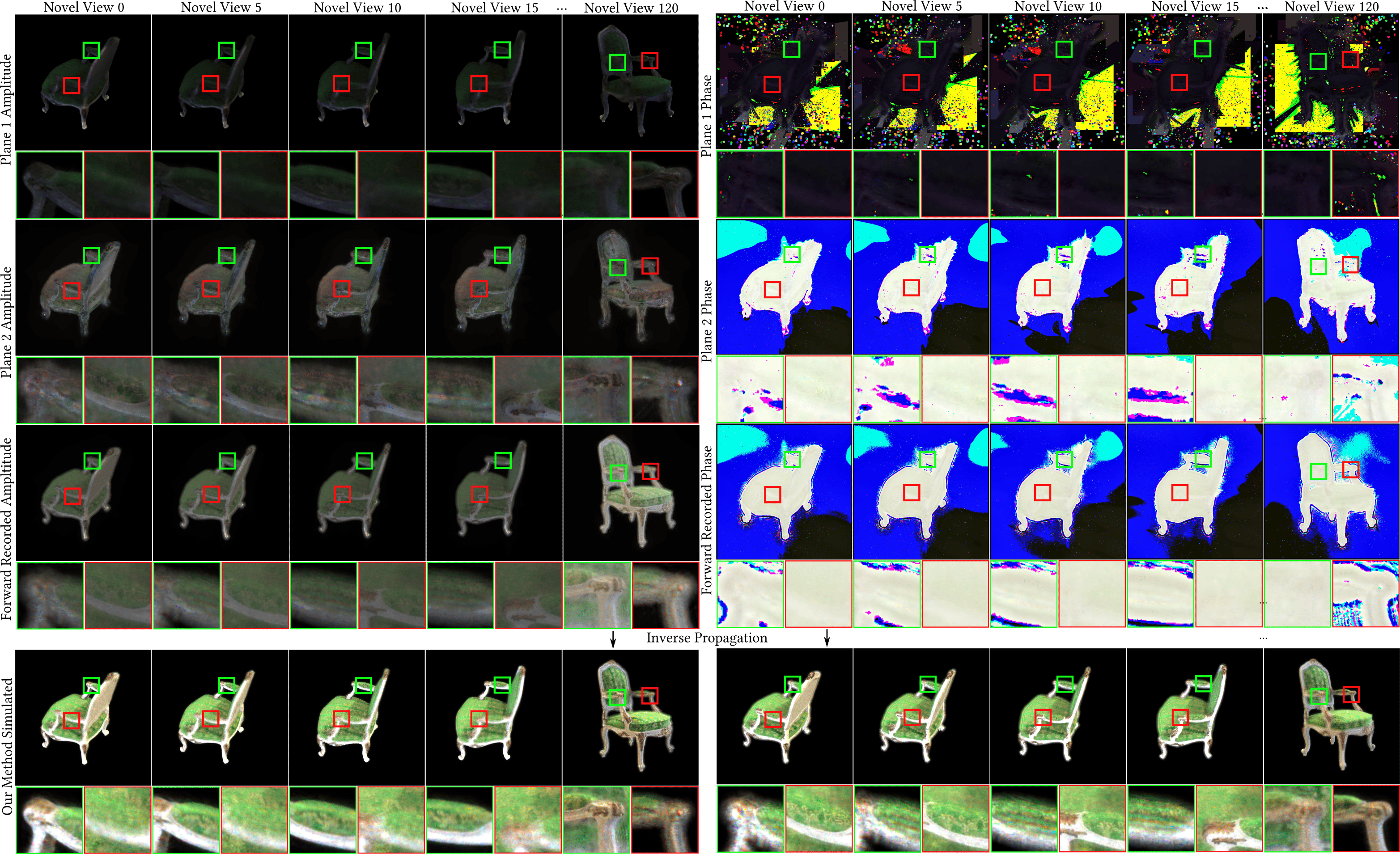}
    \caption{Gaussian primitive contributions under continuous novel-view variation.
    Columns show novel viewpoints sampled along a continuous camera trajectory.
    \textbf{Row 1:} Plane~1 complex-field projections from Trained 3D complex-valued Gaussians, visualized as amplitude (left) and phase (right).
    \textbf{Row 2:} Plane~2 complex-field projections.
    \textbf{Row 3:} Forward-recorded hologram formed by coherently combining both depth-plane fields via band-limited \ASM propagation.
    \textbf{Row 4:} Reconstructed intensities via inverse propagation, showing front- and back-focus responses.}

    \label{fig:progression}
\end{figure}
\end{landscape}

\refFig{progression} demonstrates the behavior of learned Gaussian primitives under continuous camera trajectory variation, validating our implicit geometric learning mechanism.

The first two rows show Plane~1 and Plane~2 complex-field projections (amplitude and phase). Despite using fixed plane assignment probabilities $\boldsymbol{\rho}$ learned during training,
the projected complex fields maintain correct geometric appearance across novel views. This occurs because:
\begin{enumerate}
\item \textbf{Occlusion handling through alpha blending}: Gaussians that become occluded at novel angles are naturally attenuated by the transmittance product $\prod_{j=1}^{n-1}(1-\alpha_j)$.
\item \textbf{Multi-view learned assignments}: The plane probabilities $\boldsymbol{\rho}_n$ are optimized jointly over the full training viewpoint distribution,
enabling them to function collectively rather than being viewpoint-specific.
\item \textbf{Geometric awareness}: While no Gaussian is explicitly supervised with a ground-truth 3D depth during forward recording, their collective behavior—
through learned $\boldsymbol{\rho}$ and $\alpha$ parameters—produces geometrically consistent results, which empirically generalize well to continuously interpolated novel views.
\end{enumerate}

To further clarify this behavior, consider a single Gaussian primitive in our representation. Each Gaussian is parameterized by both a plane assignment probability
$\boldsymbol{\rho}$ and an opacity $\alpha$ (Eq.~9). These parameters are learned through multi-view reconstruction supervision and are fixed after training,
making the Gaussian intrinsically view-independent and directly reusable for arbitrary novel viewpoints. While prior methods do not introduce an explicit plane assignment parameter $\boldsymbol{\rho}$,
the role of $\alpha$ in our formulation is consistent with recent work~\cite{choi2025gaussian}, where opacity similarly governs occlusion and contribution strength.

The final column (Novel View~120) demonstrates robustness under large rotation angles, where the camera viewpoint transitions from one side of the chair to the opposite side.
At this extreme angle, many Gaussians that contributed significantly at earlier views become heavily occluded; nevertheless, the reconstruction remains correct because other
Gaussians that were previously attenuated now contribute more strongly. The zoomed insets reveal that individual Gaussians maintain consistent intrinsic properties
(amplitude and phase) across viewpoints, while their effective contributions vary naturally due to occlusion.
This validates that our method achieves view-consistent holographic encoding through learned scene parameters rather than per-view recalculation.
The forward-recorded hologram (Row~3) and reconstructed intensities (Row~4)
further confirm that fixed plane assignments, combined with proper occlusion modeling, enable correct rendering throughout the camera trajectory.

\section{Extra Simulated And Captured Result}
\label{supplementary:extra_result}
\begin{figure*}[ht!]
   \centering
   \includegraphics[width=0.85\textwidth]{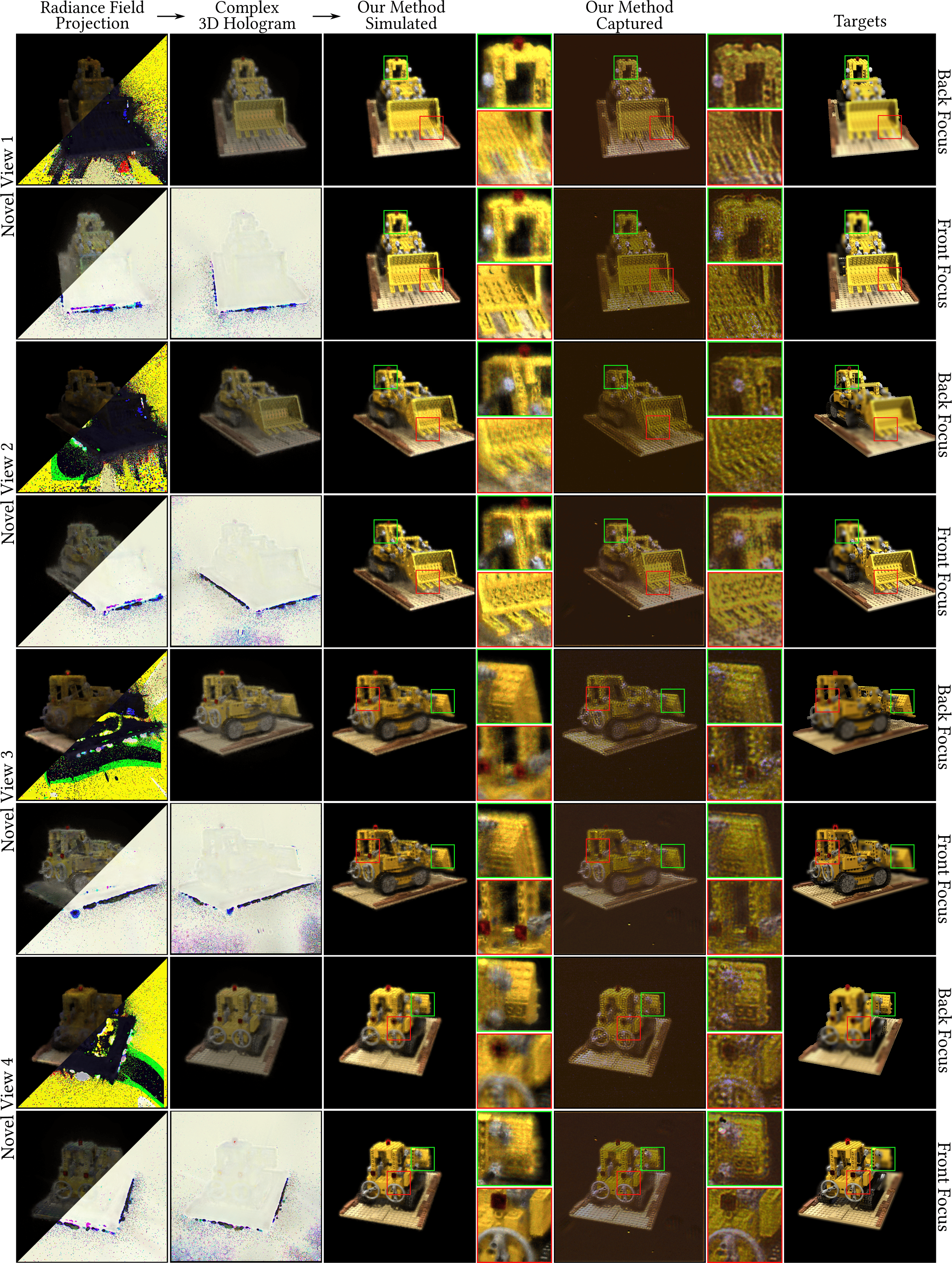}
   \caption{Extra novel-view comparison of our method on lego scene from the NeRF Synthetic dataset.
   The first two columns show the radiance field projections and their rendered complex 3D holograms.
   The central columns present our method`s simulated results and experimentally captured results.
   The rightmost column displays the target images used as optimization objectives in our method.}
   \label{fig:result_synthetic_lego}
\end{figure*}

\begin{figure*}[ht!]
   \centering
   \includegraphics[width=0.85\textwidth]{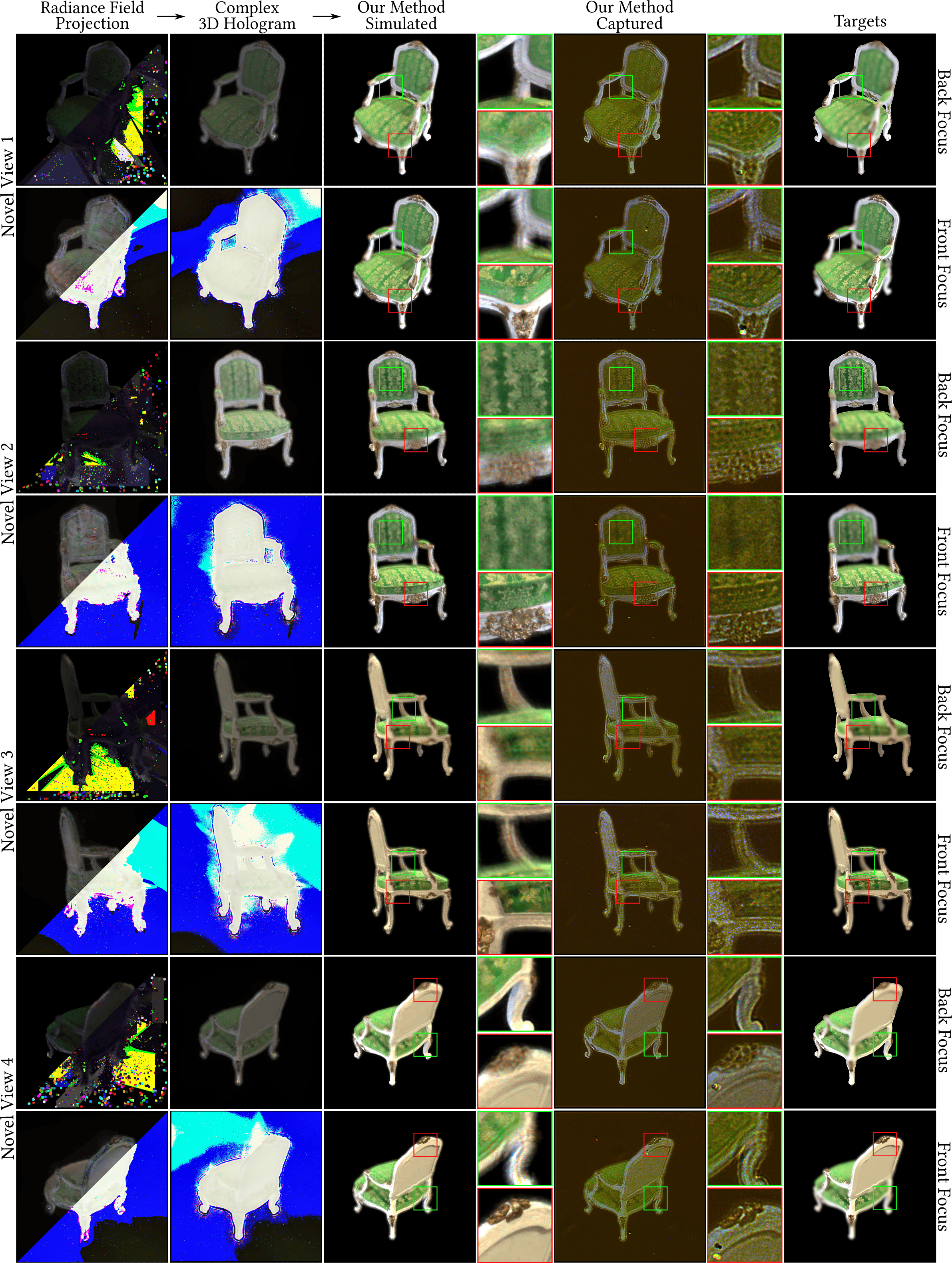}
   \caption{Extra novel-view comparison of our method on chair scene from the NeRF Synthetic dataset.
   The first two columns show the radiance field projections and their rendered complex 3D holograms.
   The central columns present our method`s simulated results and experimentally captured results.
   The rightmost column displays the target images used as optimization objectives in our method.}
   \label{fig:result_synthetic_chair}
\end{figure*}

\begin{figure*}[ht!]
   \centering
   \includegraphics[width=0.85\textwidth]{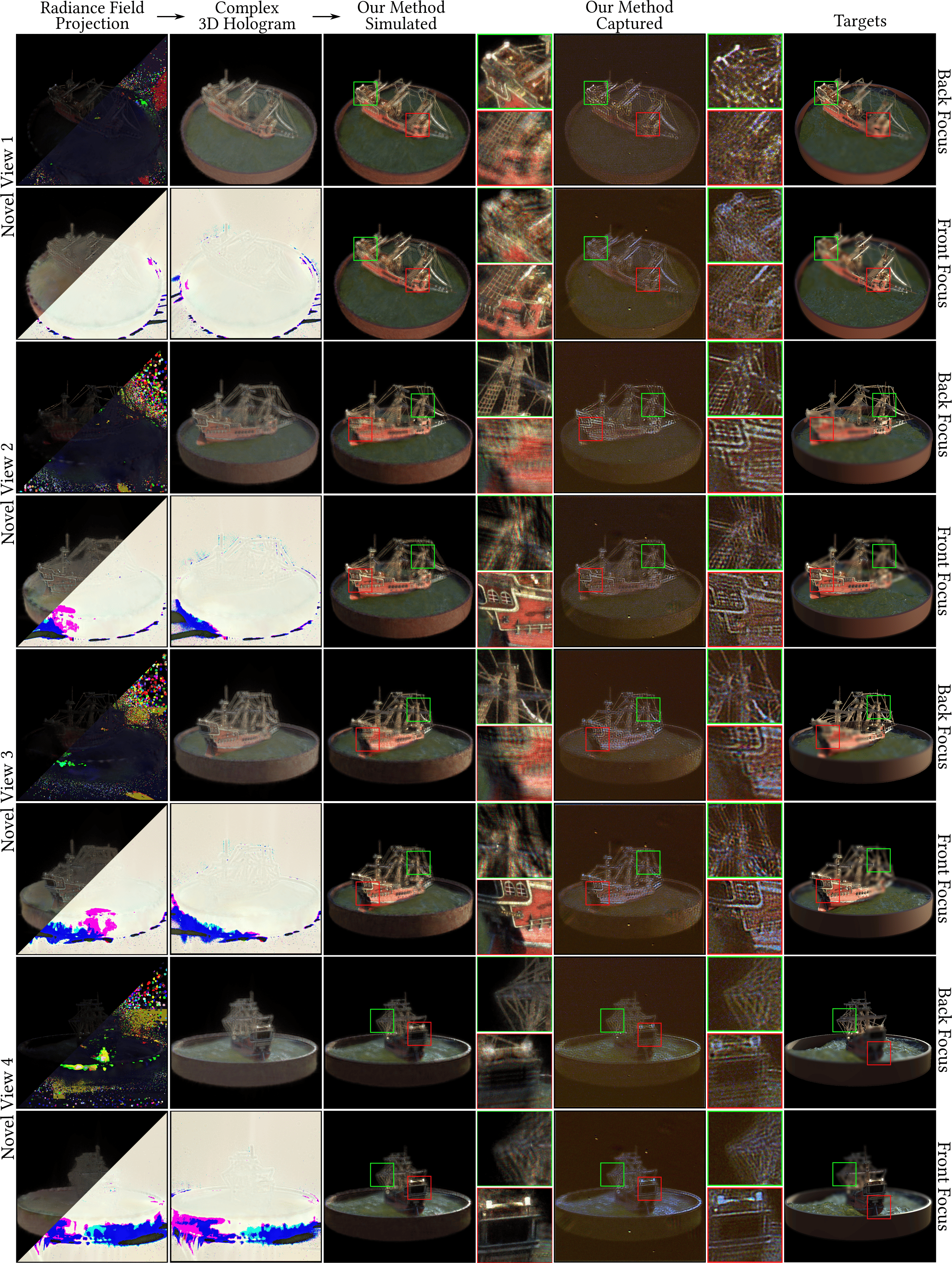}
   \caption{Extra novel-view comparison of our method on ship scene from the NeRF Synthetic dataset.
   The first two columns show the radiance field projections and their rendered complex 3D holograms.
   The central columns present our method`s simulated results and experimentally captured results.
   The rightmost column displays the target images used as optimization objectives in our method.}
   \label{fig:result_synthetic_ship}
\end{figure*}

\begin{figure*}[ht!]
   \centering
   \includegraphics[width=0.85\textwidth]{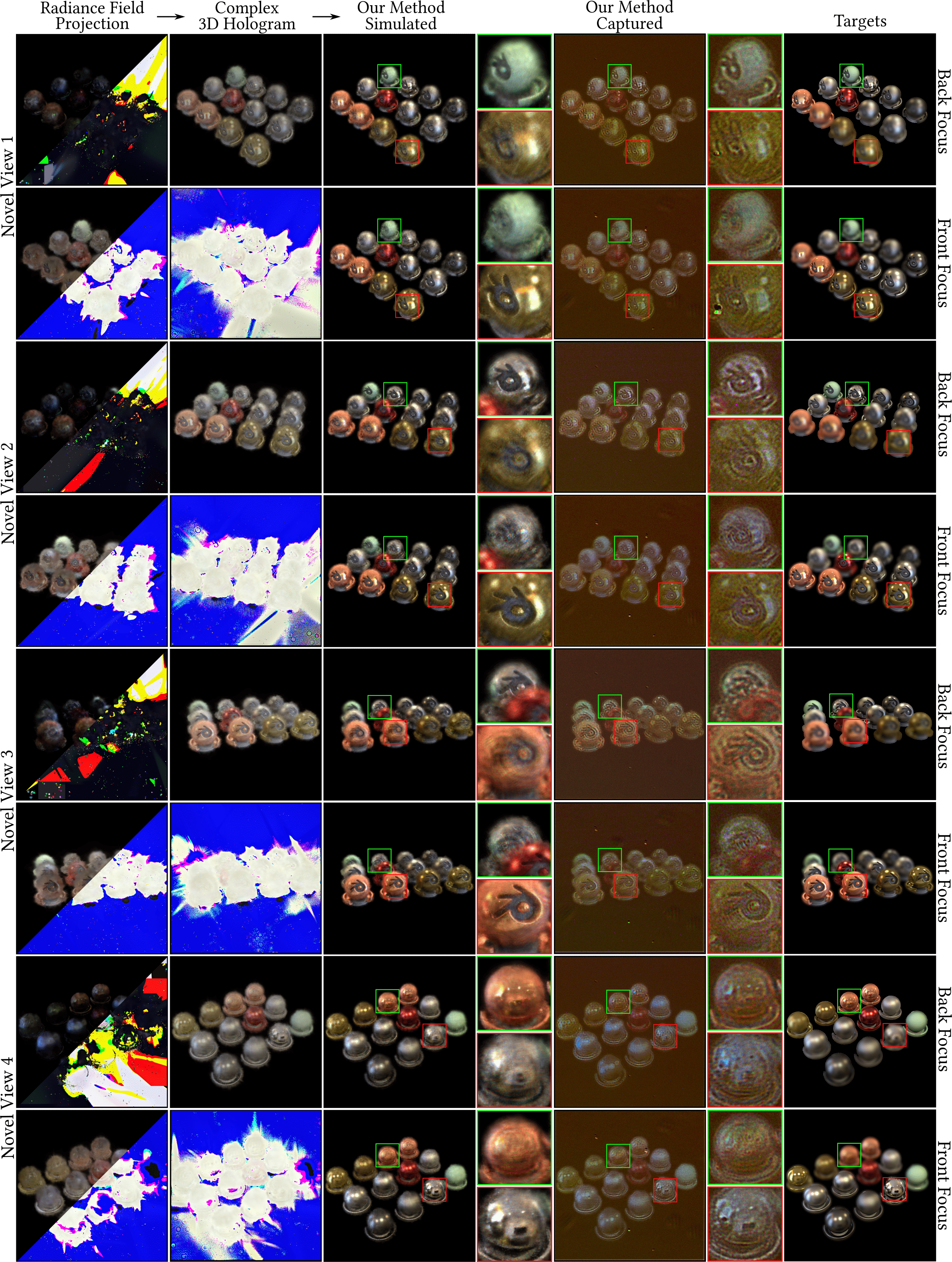}
   \caption{Extra novel-view comparison of our method on materials scene from the NeRF Synthetic dataset.
   The first two columns show the radiance field projections and their rendered complex 3D holograms.
   The central columns present our method`s simulated results and experimentally captured results.
   The rightmost column displays the target images used as optimization objectives in our method.}
   \label{fig:result_synthetic_materials}
\end{figure*}

\begin{figure*}[ht!]
   \centering
   \includegraphics[width=0.85\textwidth]{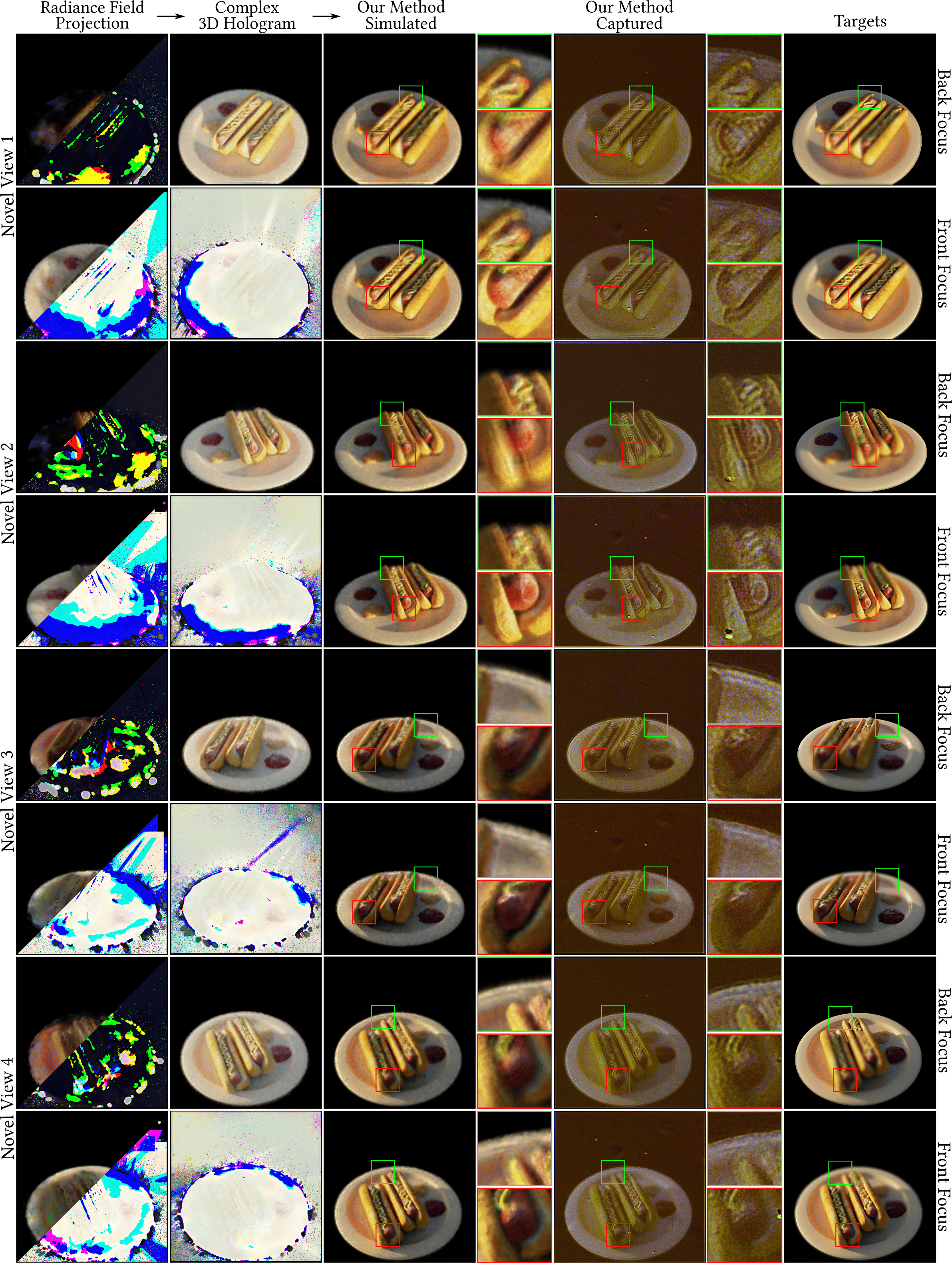}
   \caption{Extra novel-view comparison of our method on hotdog scene from the NeRF Synthetic dataset.
   The first two columns show the radiance field projections and their rendered complex 3D holograms.
   The central columns present our method`s simulated results and experimentally captured results.
   The rightmost column displays the target images used as optimization objectives in our method.}
   \label{fig:result_synthetic_hotdog}
\end{figure*}

\begin{figure*}[ht!]
   \centering
   \includegraphics[width=0.98\textwidth]{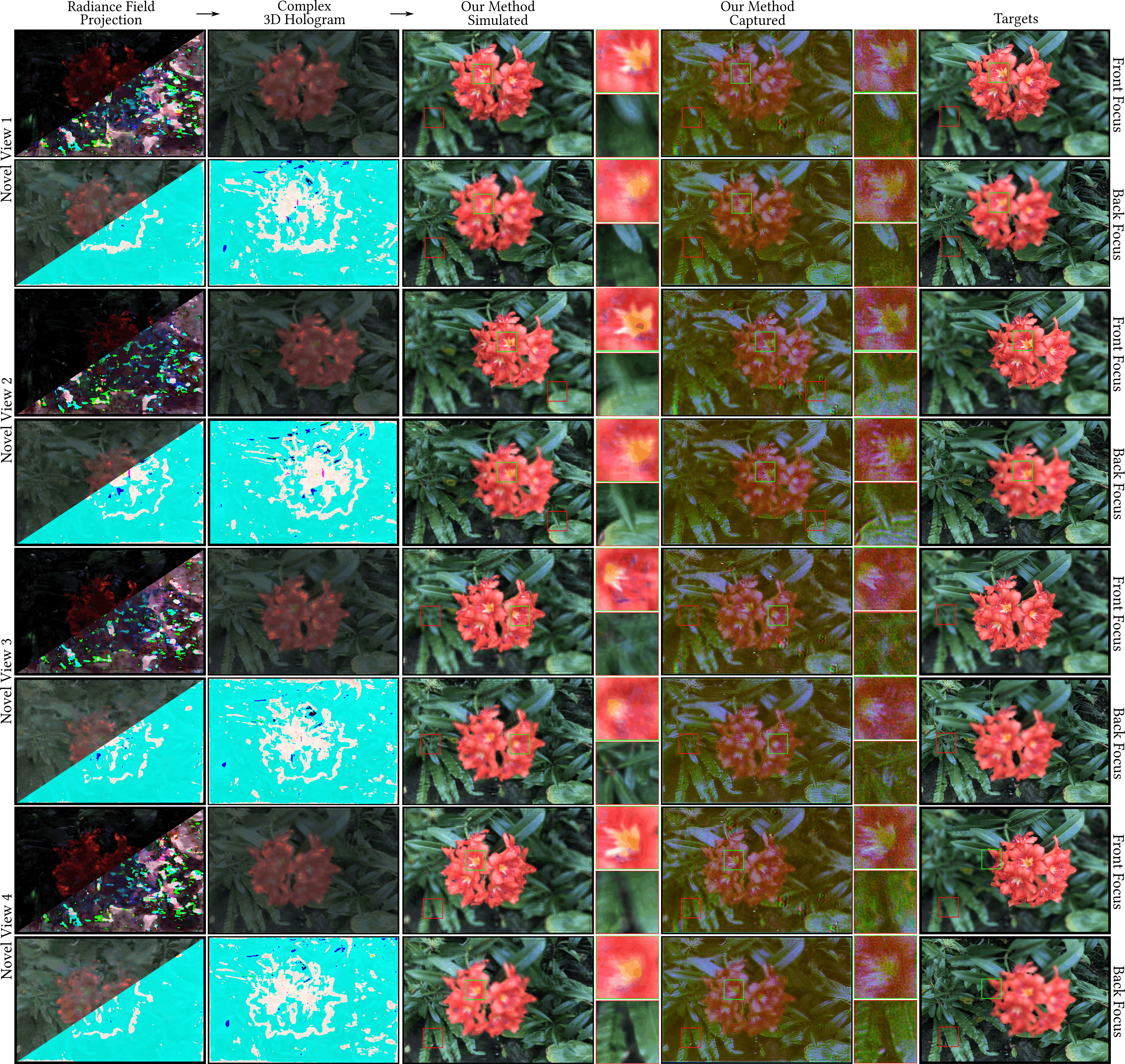}
   \caption{Extra novel-view comparison of our method on flower scene from the LLFF dataset.
   The first two columns show the radiance field projections and their rendered complex 3D holograms.
   The central columns present our method`s simulated results and experimentally captured results.
   The rightmost column displays the target images used as optimization objectives in our method.}
   \label{fig:result_flower}
\end{figure*}

\begin{figure*}[ht!]
   \centering
   \includegraphics[width=0.98\textwidth]{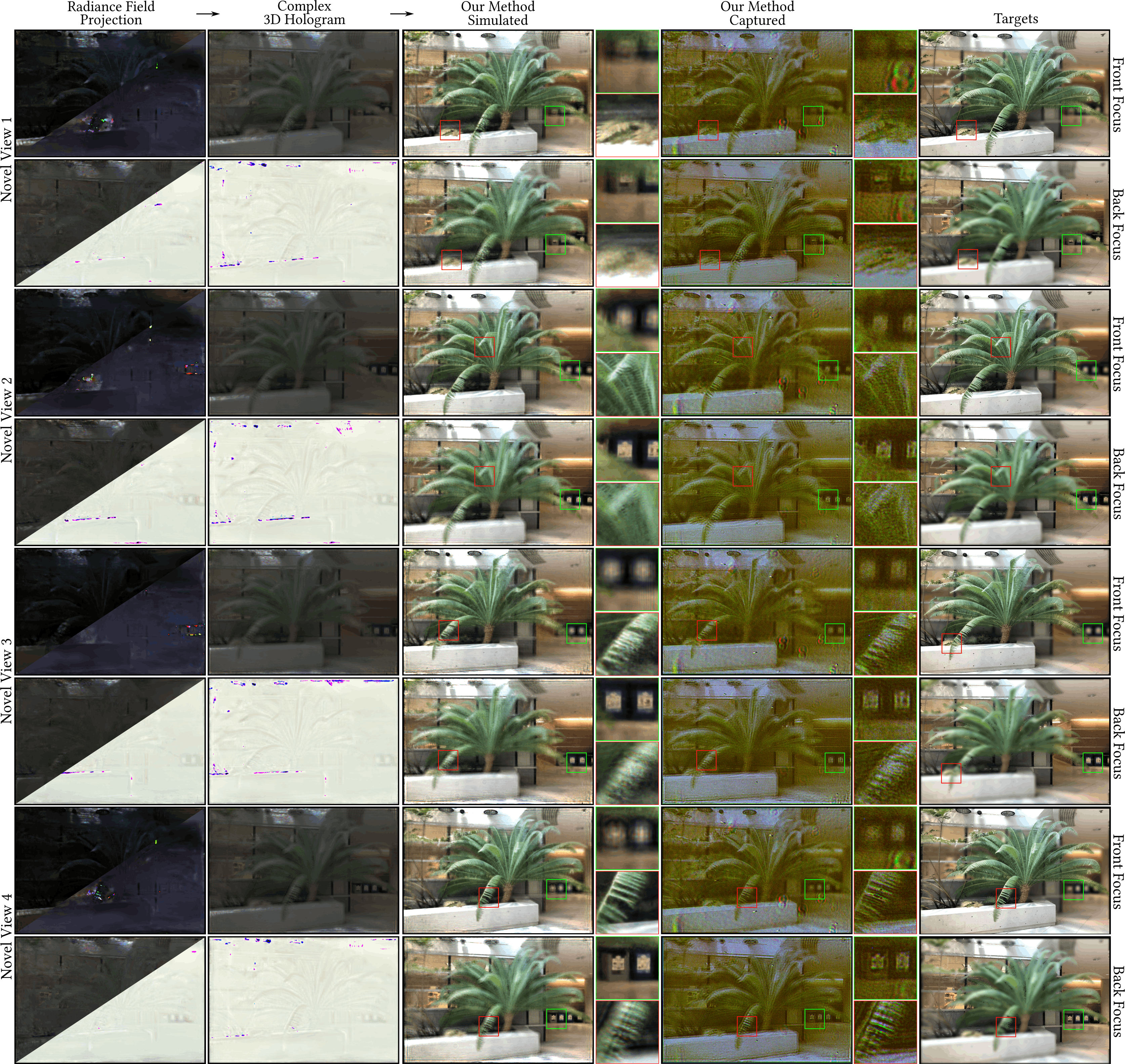}
   \caption{Extra novel-view comparison of our method on fern scene from the LLFF dataset.
   The first two columns show the radiance field projections and their rendered complex 3D holograms.
   The central columns present our method`s simulated results and experimentally captured results.
   The rightmost column displays the target images used as optimization objectives in our method.}
   \label{fig:result_fern}
\end{figure*}

\begin{figure*}[ht!]
   \centering
   \includegraphics[width=0.98\textwidth]{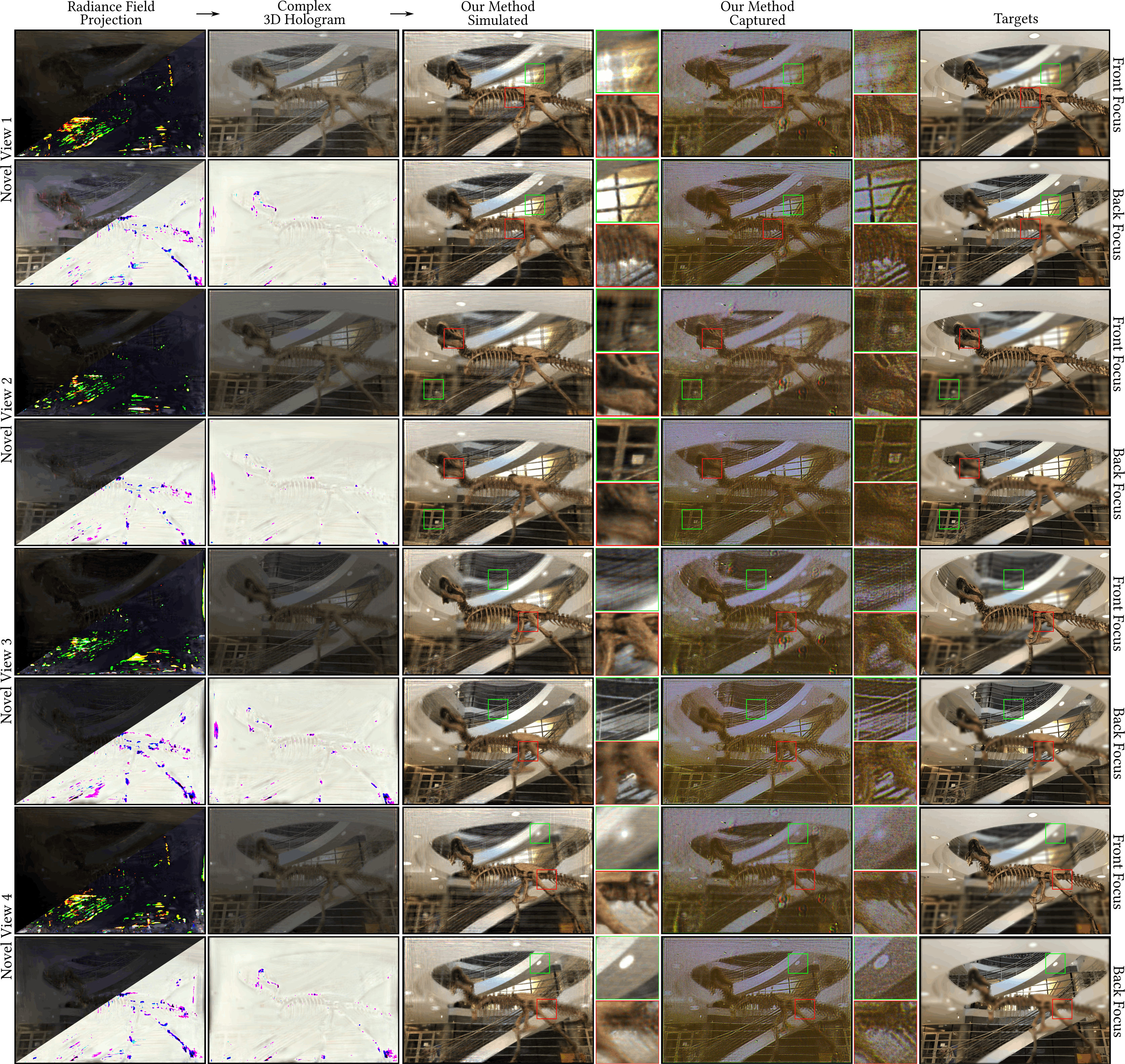}
   \caption{Extra novel-view comparison of our method on trex scene from the LLFF dataset.
   The first two columns show the radiance field projections and their rendered complex 3D holograms.
   The central columns present our method`s simulated results and experimentally captured results.
   The rightmost column displays the target images used as optimization objectives in our method.}
   \label{fig:result_trex}
\end{figure*}

\begin{figure*}[ht!]
   \centering
   \includegraphics[width=0.98\textwidth]{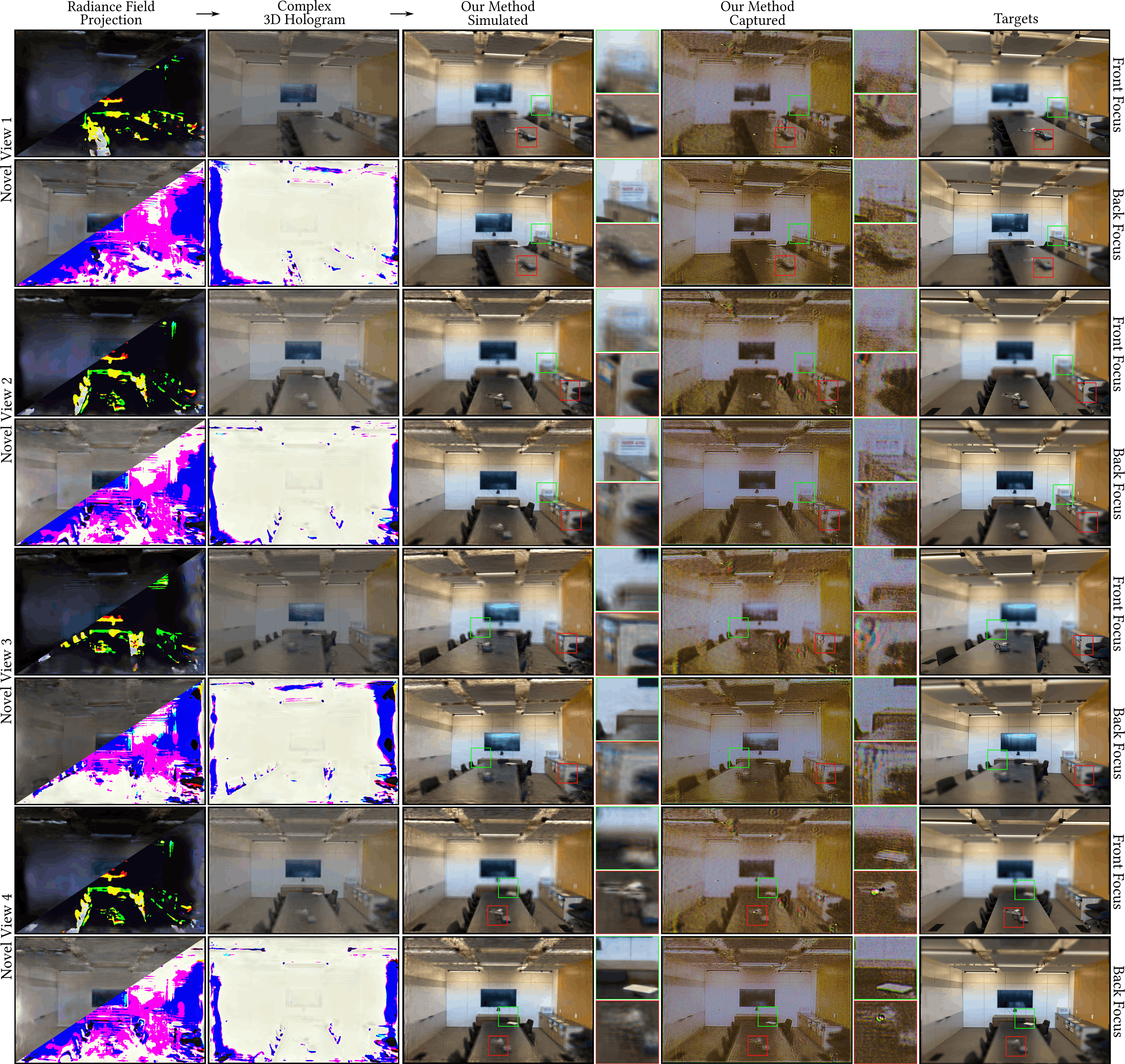}
   \caption{Extra novel-view comparison of our method on room scene from the LLFF dataset.
   The first two columns show the radiance field projections and their rendered complex 3D holograms.
   The central columns present our method`s simulated results and experimentally captured results.
   The rightmost column displays the target images used as optimization objectives in our method.}
   \label{fig:result_room}
\end{figure*}

This section presents additional experimental results demonstrating the effectiveness of our complex-valued 3D Gaussian splatting
approach across diverse scenes. We provide comprehensive comparisons on both synthetic and real-world datasets,
including scenes from the NeRF Synthetic dataset (lego in \refFig{result_synthetic_lego}, chair in \refFig{result_synthetic_chair},
ship in \refFig{result_synthetic_ship}, materials in \refFig{result_synthetic_materials}, and hotdog in \refFig{result_synthetic_hotdog})
and the LLFF dataset (flower in \refFig{result_flower}, fern in \refFig{result_fern}, trex in \refFig{result_trex},
and room in \refFig{result_room}). Each figure showcases the complete pipeline from radiance field projections and
complex 3D hologram rendering to both simulated and experimentally captured results,
demonstrating consistent quality across different scene types and lighting conditions.


\end{document}